\documentclass[reprint,amsmath,amsfonts,amssymb,preprintnumbers]{revtex4-2}

\usepackage[utf8]{inputenc}
\usepackage[T1]{fontenc}
\usepackage{lmodern}
\usepackage{graphicx}
\usepackage{dcolumn}
\usepackage{bm}
\usepackage{textcomp}
\usepackage{ulem}
\usepackage{ifpdf}
\usepackage[squaren,Gray]{SIunits}
\usepackage{color}

\definecolor{red}{rgb}{1,0,0}
\definecolor{blue}{rgb}{0,0,1}
\definecolor{darkred}{rgb}{0.6,0,0}
\definecolor{darkblue}{rgb}{0,0,0.6}
\definecolor{darkgreen}{rgb}{0,0.5,0}
\definecolor{grey}{rgb}{0.5,0.5,0.5}

\ifpdf
\usepackage{epstopdf}
\usepackage[pdftex,unicode,pdfstartview={FitH},pdfborder={0 0 0}]{hyperref}
\usepackage{hypcap}
\else
\usepackage[hypertex]{hyperref}
\fi

\hypersetup{
    bookmarksnumbered = true,
    colorlinks = true, linkcolor = black,
    citecolor = black, filecolor = black,
    menucolor = black, urlcolor = black
}

\newcolumntype{R}{>{$\displaystyle}r<{$}}
\newcolumntype{C}{>{$\displaystyle}c<{$}}

\begin{document}

\title{Lattice reconstruction in MoSe$_2$-WSe$_2$ heterobilayers\\
synthesized by chemical vapor deposition}


\author{Zhijie Li$^{1,\ddag}$, Farsane Tabataba-Vakili$^{1,2,*,\ddag}$, Shen Zhao$^{1,\ddag}$, Anna Rupp$^{1}$, Ismail Bilgin$^{1}$, Ziria Herdegen$^{3}$, Benjamin M\"arz$^{3}$, Kenji Watanabe$^{4}$, Takashi Taniguchi$^{5}$, Gabriel Ravanhani Schleder$^{6}$, Anvar~S.~Baimuratov$^{1}$, Efthimios Kaxiras$^{6,7}$, Knut M\"uller-Caspary$^{3}$, and Alexander H\"ogele$^{1,2,*}$\footnote[0]{$^{*}$ Corresponding authors: F.~T.-V. (f.tabataba@lmu.de) and A. H. (alexander.hoegele@lmu.de).}\footnote[0]{$^{\ddag}$ Z.~L., F.~T.-V. and S.~Z. contributed equally to this work.}}

\affiliation{$^1$Fakult\"at f\"ur Physik, Munich Quantum Center, and Center for NanoScience (CeNS), Ludwig-Maximilians-Universit\"at M\"unchen, Geschwister-Scholl-Platz 1, 80539 M\"unchen, Germany}

\affiliation{$^2$Munich Center for Quantum Science and Technology (MCQST), Schellingtra\ss{}e 4, 80799 M\"unchen, Germany}

\affiliation{$^3$Department of Chemistry and Center for NanoScience, Ludwig-Maximilians-Universit\"at M\"unchen, Butenandtstr. 11, 81377 M\"unchen, Germany}

\affiliation{$^4$Research Center for Functional Materials, National Institute for Materials Science, 1-1 Namiki, Tsukuba 305-0044, Japan}

\affiliation{$^5$International Center for Materials Nanoarchitectonics, National Institute for Materials Science, 1-1 Namiki, Tsukuba 305-0044, Japan}

\affiliation{$^6$John A. Paulson School of Engineering and Applied Sciences, Harvard University, Cambridge, Massachusetts 02138, USA}

\affiliation{$^7$Department of Physics, Harvard University, Cambridge, Massachusetts 02138, USA}

\affiliation{$^*$Corresponding authors: F.~T.-V. (f.tabataba@lmu.de) and A. H. (alexander.hoegele@lmu.de).}

\affiliation{$^{\ddag}$Z.~L., F.~T.-V. and S.~Z. contributed equally to this work.}


\begin{abstract}
Vertical van der Waals heterostructures of semiconducting transition metal dichalcogenides realize moir\'e systems with rich correlated electron phases and moir\'e exciton phenomena. For material combinations with small lattice mismatch and twist angles as in MoSe$_2$-WSe$_2$, however, lattice reconstruction eliminates the canonical moir\'e pattern and instead gives rise to arrays of periodically reconstructed nanoscale domains and mesoscopically extended areas of one atomic registry. Here, we elucidate the role of atomic reconstruction in MoSe$_2$-WSe$_2$ heterostructures synthesized by chemical vapor deposition. With complementary imaging down to the atomic scale, simulations, and optical spectroscopy methods we identify the coexistence of moir\'e-type cores and extended moir\'e-free regions in heterostacks with parallel and antiparallel alignment. Our work highlights the potential of chemical vapor deposition for applications requiring laterally extended heterosystems of one atomic registry or exciton-confining heterostack arrays.
\end{abstract}

\maketitle


Vertical heterostructures of transition metal dichalcogenide semiconductors manifest in two contrasting regimes. On the one hand, exfoliation-stacked heterobilayers with finite lattice mismatch or twist angle give rise to periodic two-dimensional (2D) moir\'e patterns, which in turn result in flat moir\'e minibands of charge carriers with rich phenomena of correlated Hubbard model physics \cite{wu2018hubbard,tang2020simulation,shimazaki2020strongly, regan2020mott,xu2020correlated,tang2022dielectric}. Moir\'e potentials also profoundly affect strongly bound electron-hole pairs in the form of intralayer \cite{zhang2018moire,naik2022intralayer} and interlayer \cite{seyler2019signatures,tran2019evidence,jin2019observation,alexeev2019resonantly} excitons formed by Coulomb correlations within or across individual layers. This scenario is contrasted by moir\'e-free heterobilayers on the other hand, obtained from chemical vapor deposition (CVD) synthesis \cite{hsu2018negative,choi2020moire,xia2021strong}, where the absence of lateral moir\'e potentials is signified by one atomic registry extending laterally over large sample areas \cite{xia2021strong} with simple photoluminescence (PL) spectra \cite{hsu2018negative} or enhanced diffusion of interlayer excitons \cite{choi2020moire}. 

Heterostructure systems with diffusive interlayer excitons represent an ideal material platform for integrated dipolar exciton circuits \cite{Butov2017}, as demonstrated recently in exfoliation-stacked heterostructures with an additional interfacial layer of hexagonal boron nitride (hBN), which mitigates the exciton-confining moir\'e potential \cite{Sun2022,Shanks2022}. Even more promising are CVD-synthesized heterobilayers free of moir\'e effects, featuring enhanced diffusivity of dipolar interlayer excitons \cite{choi2020moire}, unaffected by diffusion-inhibiting moir\'e confinement \cite{choi2020moire,Yuan2020,Wang2021}. Such moir\'e-free heterostructures are not only ideal for integrated exciton circuits with external control by electrostatic gates \cite{Jiang2021,Ciarrocchi2022}, they could also enable deterministically imprinting arbitrary, tunable potential landscapes via patterned gate-electrodes \cite{shanks2021nanoscale,shanks2022single} or dielectric superlattices \cite{shi2019gate}. 

\begin{figure*}[t!]
\includegraphics[scale=1.02]{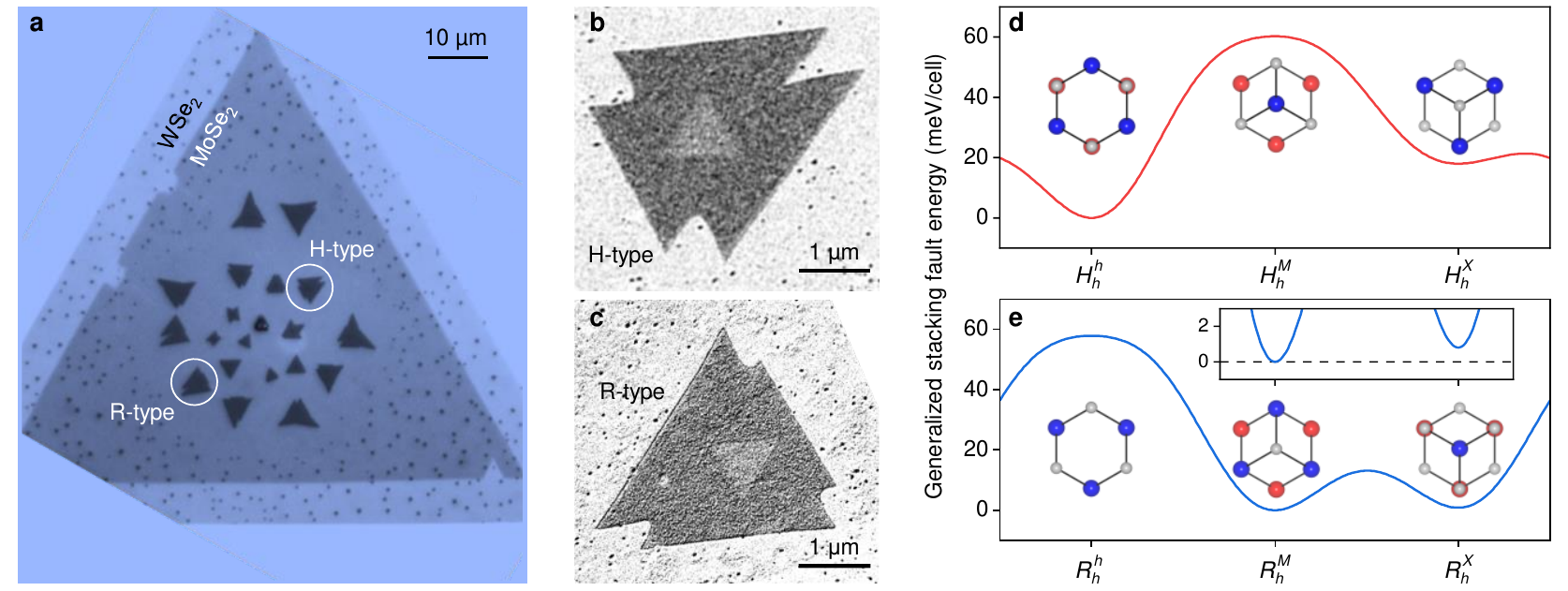}
\caption{\textbf{MoSe$_2$-WSe$_2$ heterobilayers: sample structure of CVD-synthesized H- and R-type stacks and theory of stacking fault energy.} \textbf{a} Optical micrograph of an as-grown heterobilayer consisting of a large triangular MoSe$_2$ monolayer with an outer edge of monolayer WSe$_2$ and triangles of monolayer WSe$_2$ on top with 0° (R-type) and 180° (H-type) twist. The white circles indicate the H- and R-type flakes studied by optical spectroscopy. \textbf{b}, \textbf{c} SEM images (recorded with secondary electron imaging and shown with inverted black-and-white contrast) of two representative H- and R-type heterobilayers synthesized in the same growth. \textbf{d}, \textbf{e} DFT calculations of the generalized stacking fault energy for the different stackings in H- and R-type heterostructures, respectively, including top views of the three high-symmetry atomic registries with  W (blue, top layer), Mo (red, bottom layer),  and Se (gray) atoms. Inset in \textbf{e}: Zoom to the minima at $R_h^M$ and $R_h^X$ stackings.} \label{fig1}
\end{figure*}

Moir\'e-free domains on micron length scales also emerge in heterostructures with small lattice mismatch and marginal twist subject to mesoscopic lattice reconstruction, where the driving mechanism behind atom rearrangement into energetically favorable registries is provided by the competition between intralayer strain and interlayer adhesion energy \cite{carr2018relaxation,enaldiev2020stacking,rosenberger2020twist}, yielding mesoscopic 2D domains of only one registry in MoSe$_2$-WSe$_2$ stamping-assembled heterostacks \cite{zhao2022excitons}. For practical applications, however, such non-deterministic fabrication methods with resulting spatial inhomogeneities in morphology and optical properties \cite{zhao2022excitons} limit the required uniformity and scalability, rendering CVD-based approaches to large-area moir\'e-free systems a promising alternative.

In this work, we present an elaborate study of CVD-synthesized vertical MoSe$_2$-WSe$_2$ heterobilayers with evidence for extended reconstruction into domains of one atomic registry, enclosing a central region of periodically reconstructed nanoscale domains \cite{zhao2022excitons,carr2018relaxation,enaldiev2020stacking}. Our studies cover both high-symmetry stacking configurations with 0° (R-type) and 180° (H-type) twist angle, and employ complementary imaging and optical spectroscopy methods to identify the diversity in local configurations of the reconstructed crystal lattice and the respective signatures of exciton transitions. Our work highlights the potential of CVD synthesis for obtaining both extended moir\'e-free domains and exciton-confining arrays of periodically reconstructed moir\'e regions, realizing the limits of dipolar excitons with and without a spatially varying potential landscape.


\begin{figure*}[t!]
\includegraphics[scale=1.0]{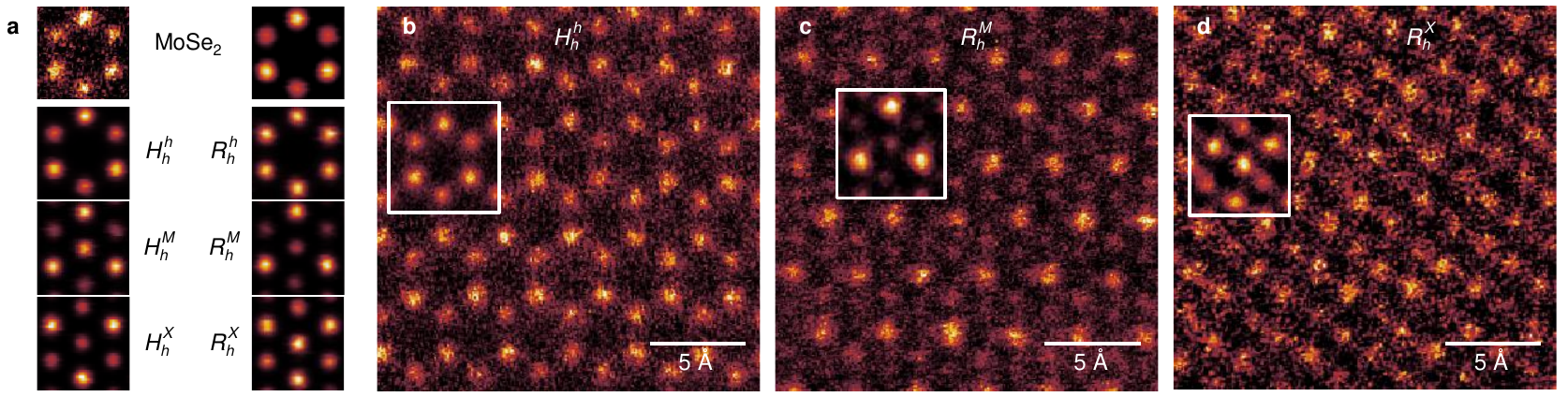}
\caption{\textbf{HAADF-STEM imaging and simulations.} \textbf{a}, Top row: Measurement (left) and simulation (right) of MoSe$_2$ monolayer. Bottom rows: Simulations of the three high-symmetry stackings in H-type (left column) and R-type (right column) heterostacks. \textbf{b - d}, HAADF-STEM images of heterobilayers in $H_h^h$, $R_h^M$, and $R_h^X$ stacking, respectively. Insets show images averaged over ten unit cells.} \label{fig2}
\end{figure*}

Our samples with MoSe$_2$-WSe$_2$ heterobilayers consist of large monolayers of MoSe$_2$ with monolayers of WSe$_2$ on top, both synthesized by CVD in a two-step growth (see Section~1 and 2 in
the Supporting Information for details). Figure~\ref{fig1}a depicts an optical micrograph of an as-grown heterobilayer, showing the formation of both H- and R-type stackings. We examined another heterobilayer sample synthesized in the same growth run by scanning electron microscopy (SEM), with representative images shown in Fig.~\ref{fig1}b, c. The images of H- and R-type samples were recorded with secondary electron imaging with material- and stacking-sensitive contrast \cite{ashida2015crystallographic,andersen2021excitons,zhao2022excitons} (see Section~3 in the Supporting Information for details), clearly discerning a small triangle in the center of the heterostack.

Three high-symmetry atomic registries are distinguished in H- and R-type heterostructures, namely $H_h^h$, $H_h^M$, $H_h^X$ and $R_h^h$, $R_h^M$, $R_h^X$, with $M$, $X$, and $h$ referring to the transition metal, chalcogen, and center of the hexagon, and the superscript and subscript to the MoSe$_2$  and WSe$_2$ layer, respectively (note that the nomenclature is chosen to be consistent with previous work \cite{yu2017moire,wozniak2020exciton, zhao2022excitons}). In Fig.~\ref{fig1}d, e, we show density functional theory (DFT) calculations of the generalized stacking fault energy (GSFE) \cite{carr2018relaxation} for the different atomic registries with local extrema at the three high-symmetry stackings (see Section~4 in
the Supporting Information for details) indicated by the respective top-view schematics. Minima in GSFE correspond to the energetically most favorable stackings that dominate the reconstruction \cite{rosenberger2020twist}. In the H-type case shown in Fig.~\ref{fig1}d, the $H_h^h$ atomic registry is by far the energetically most favorable, consistent with prior work \cite{carr2018relaxation,zhao2022excitons,rosenberger2020twist}. In contrast, in the R-type case (Fig.~\ref{fig1}e) both $R_h^M$ and $R_h^X$ stackings are close to the optimal energy, with $R_h^M$ being slightly more favorable (inset of Fig.~\ref{fig1}e). This implies close competition between the two registries, with a higher likelihood of $R_h^M$ to form extended domains.

Our study of individual heteroflakes with aberration-corrected high-resolution high-angle annular dark-field scanning transmission electron microscopy (HAADF-STEM) (see Section~5 in
the Supporting Information for details) confirms the exclusive dominance of the $H_h^h$ atomic registry in H-type heterostacks, as well as the reconstruction of R-type stacks into $R_h^M$ and $R_h^X$ registries. To begin with, we establish the orientation of the crystallographic axes in the heterobilayers using the surrounding MoSe$_2$ monolayer, with measurement (left) and simulation (right) of the hexagonal lattice shown in the top row of Fig.~\ref{fig2}a. For this orientation, the bottom rows in Fig.~\ref{fig2}a show simulations of the three possible high-symmetry stackings in H- and R-type, which compete for energy minimization in reconstruction. In stacks of H-type, our experiments with results as in Fig.~\ref{fig2}b identify the $H_h^h$ stacking only, in agreement with the theoretical prediction above. In R-type heterostacks, on the other hand, we observe extended areas of $R_h^M$ registry in some flakes (as in Fig.~\ref{fig2}c), while other flakes exhibit extended areas of $R_h^X$ stacking (as in Fig.~\ref{fig2}d).  Given the close similarity of the two registry configurations with regard to optimal energy, it is plausible that some R-type flakes reconstruct into $R_h^M$ while others take on the $R_h^X$ stacking, and that their distribution is stochastic on the flake-to-flake case.

\begin{figure}[t!]
\includegraphics[scale=1.0]{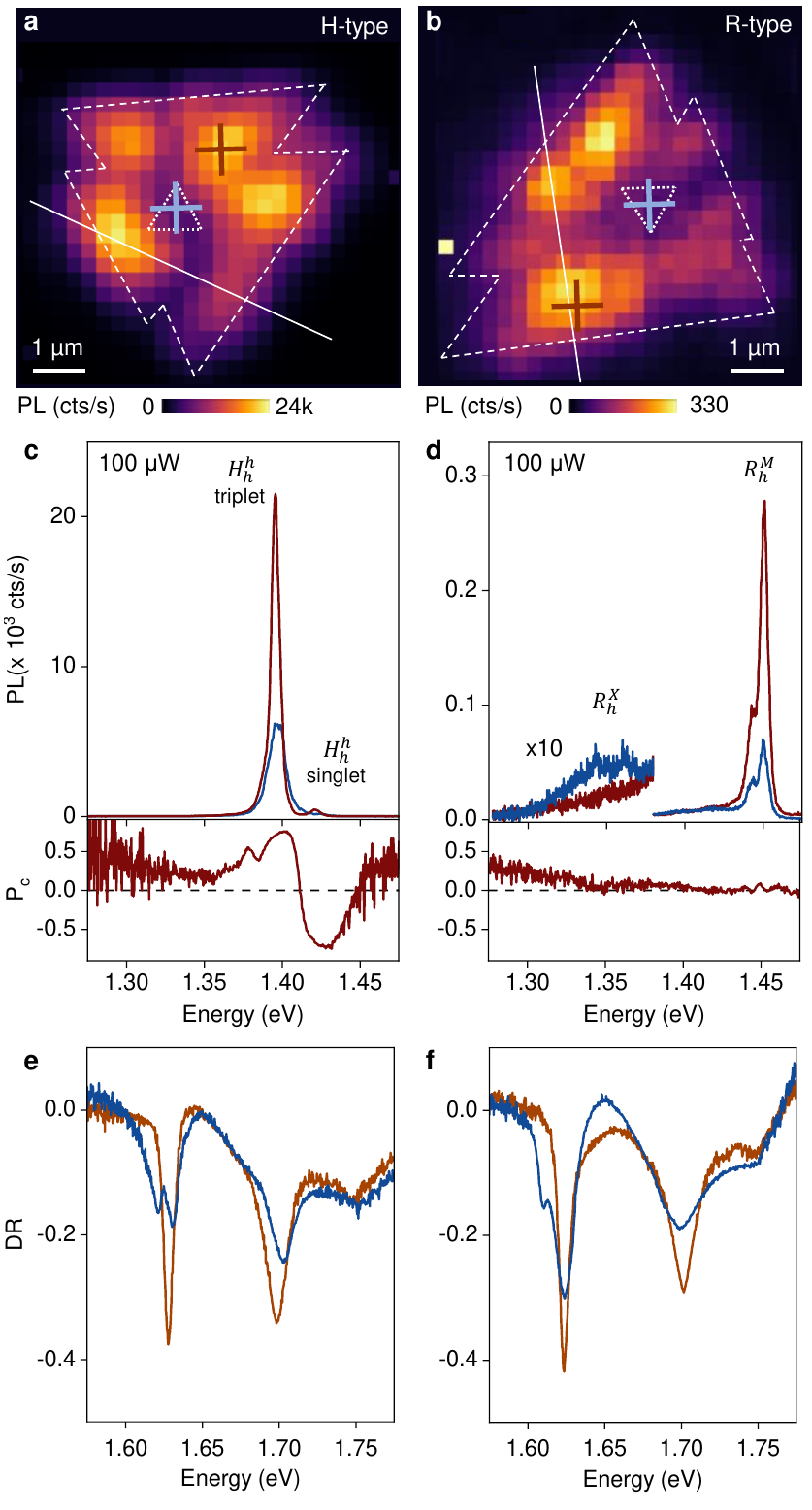}
\vspace{-8pt}
\caption{\textbf{CVD-heterobilayers in cryogenic spectroscopy.} \textbf{a}, \textbf{b} Maps of interlayer PL maximum intensity (dashed lines show flake boundaries from optical images, dotted lines delimit the central triangles, and solid lines indicate from top left to bottom right intensity profiles in Fig.~5c, d). \textbf{c}, \textbf{d} Top panel: Interlayer exciton PL spectra recorded with linearly polarized excitation and circularly polarized detection at positions shown by the brown and blue crosses in \textbf{a}, \textbf{b} (the spectra below $1.37$~eV in \textbf{d} are multiplied by $10$ for better visibility), with atomic registries assigned to the corresponding transitions. Bottom panel: Degree of circular polarization $P_\text{c}$ measured at the position of the brown cross in \textbf{a}, \textbf{b}. \textbf{e}, \textbf{f} Differential reflectance (DR) spectra corresponding to areas indicated by brown and blue crosses in \textbf{a}, \textbf{b}. All data recorded at $4$~K, with $100~\mu$W excitation in PL spectroscopy.} 
\label{fig3}
\end{figure}

To study the optical properties of CVD-synthesized H- and R-type crystals, we encapsulated the sample shown in Fig.~\ref{fig1}a in hBN (see Section~6 in the Supporting Information for details) and performed cryogenic confocal PL and differential reflectance (DR) spectroscopy. Each atomic registry uniquely determines the combination of transition energies \cite{yu2017moire,wu2018theory,zhao2022excitons}, optical selection rules \cite{yu2017moire,wu2018theory,forg2019cavity}, and oscillator strengths \cite{gillen2018interlayer,forg2021moire} of interlayer excitons, and thus provides means for spectroscopic characterization. For our sample, the maps of maximum interlayer exciton PL intensity in H- and R-type heterostructures are shown in Fig.~\ref{fig3}a, b with dashed lines indicating the heterostack boundaries. The H-type sample shows extended areas of bright interlayer exciton PL (as on the spot marked by the brown cross in Fig.~\ref{fig3}a) and a dark region near the center (blue cross in Fig.~\ref{fig3}a). The R-type sample features similar local variations in the PL map, yet with nearly two orders of magnitude lower intensity. 

The low intensity of R-type as compared to H-type stacks is surprising, as reconstructed mechanically stacked samples support the opposite trend \cite{zhao2022excitons}. This observation suggests that domains of $R_h^M$ registry with relatively dark interlayer excitons dominate CVD-grown R-type heterostacks. In the areas of maximum PL intensity marked by brown crosses in Fig.~\ref{fig3}a and b, both H- and R-stacks feature simple spectra with dominant peaks at $1.40$ and $1.45$~eV in Fig.~\ref{fig3}c and d, respectively. In the H-stack, the PL peak at $1.40$~eV with a positive degree of circular polarization $P_\text{c}$ (shown in the bottom panel of Fig.~\ref{fig3}c) corresponds to the triplet interlayer exciton transition in extended $H_h^h$ domains of reconstructed mechanically stacked samples \cite{zhao2022excitons}, and is accompanied by a weak hot-luminescence singlet at $1.42$~eV with negative $P_\text{c}$ \cite{hsu2018negative,hanbicki2018double,joe2021electrically,zhao2022excitons}.
The main peak in the R-type stack, on the other hand, is blue-shifted by $120$~meV from its counterpart observed on reconstructed $R_h^X$ domains of mechanically stacked heterostructures at $1.33$~eV \cite{zhao2022excitons} and exhibits zero $P_\text{c}$ in Fig.~\ref{fig3}d. This finding, combined with the PL energy position and reduced relative brightness, corroborate our assumption about the $R_h^M$ character of interlayer exciton states with $z$-polarized transitions \cite{forg2019cavity,forg2021moire,zhao2022excitons}. Finally, we observe that areas of maximum PL intensity exhibit simple resonances of intralayer exciton in the DR spectra of H- and R-type stacks (shown in Fig.~\ref{fig3}e, f for positions indicated by brown crosses in Fig.~\ref{fig3}a, b) as a hallmark of extended areas of one dominant atomic registry \cite{zhao2022excitons}. We also point out the absence of trion-related resonances in the DR spectra of both stacks as a signature of low residual doping, which we also confirmed with PL spectroscopy away from heterostacks on monolayer MoSe$_2$ (data not shown).

In the central areas of H- and R-type stacks marked by blue crosses in the areas of dotted triangles in Fig.~\ref{fig3}a, b and spectra in Fig.~\ref{fig3}c, d, we observe reduced intensity of the main PL peaks and an additional spectral feature around $1.35$~eV with negative $P_\text{c}$ (not shown) in the R-type case. The latter is a feature of $R_h^X$ atomic registry \cite{zhao2022excitons}, which coexists with $R_h^M$ in the core of the heterostack. This conclusion, substantiated by the observation of doublets in the DR spectra of Fig.~\ref{fig3}e, f near the intralayer exciton resonance of MoSe$_2$ at $1.62$~eV, identifies the central triangles as arrays of reconstructed quasi zero-dimensional (0D) domains \cite{zhao2022excitons}, which can form as nanoscale hexagons of $H_h^h$ registry in H-type and in triangular tiling of $R_h^X$ and $R_h^M$ domains in R-type stacks \cite{rosenberger2020twist}. Although the spatial resolution of the SEM in secondary-electron imaging is insufficient to detect the actual tiling geometry, the difference in the contrast between the outer regions of both types of heterostacks and their triangular cores in Fig.~\ref{fig1}b, c provides compelling support for the assignment of the triangular centers to reconstructed 0D arrays.

\begin{figure}[t!]
\includegraphics[scale=1.0]{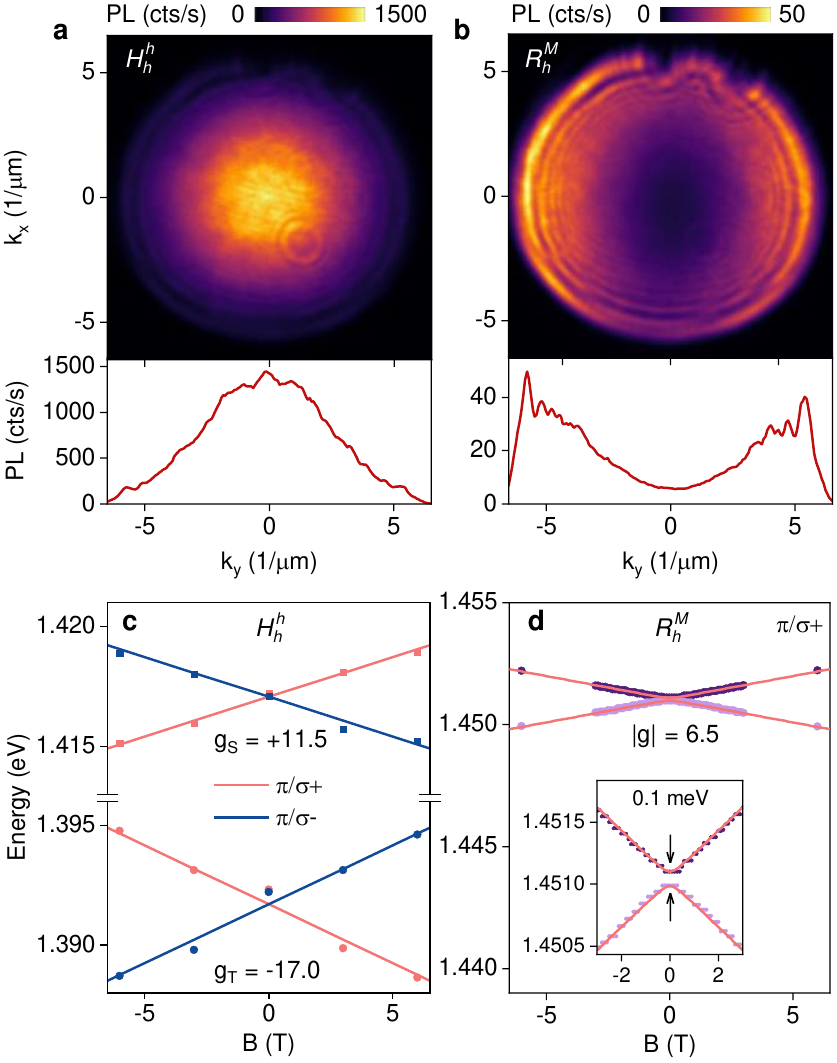}
\caption{\textbf{Momentum-space and magneto-optical characteristics of interlayer exciton luminescence.} \textbf{a}, \textbf{b} Top panels: Momentum-space maps of interlayer exciton PL in H- and R-type samples, respectively (tunable long- and short-pass filters were used to limit the detection in \textbf{b} to the energy range of $1.43 - 1.46$~eV). Bottom panels: Emission profiles at $k_x = 0$. \textbf{c}, \textbf{d} Magneto-dispersion recorded under linearly polarized excitation and circularly polarized detection, with Land\'e $g$-factors extracted from fits to the data. Inset in \textbf{d}: Zoom to the dispersion around zero field, with emphasis on the dark exciton exchange splitting $\delta$ of $0.1$~meV. All measurements were performed at $100~\mu$W excitation power and $4$~K.} \label{fig4}
\end{figure}

To elaborate on the nature of interlayer excitons in extended $R_h^M$ domains not reported previously for MoSe$_2$-WSe$_2$ heterostacks, we performed angle-resolved PL measurements. Using momentum-space imaging \cite{dominguez2014fourier} with the 2D $k$-space profile shown in Fig.~\ref{fig4}a for the PL from a bright H-type area, we first confirm the in-plane character of the optical dipole orientation for $H_h^h$ interlayer excitons for reference. Their Gaussian emission profile, with highest intensity at zero in-plane $k$-vector in the linecut at $k_x=0$ (bottom panel of Fig.~\ref{fig4}a) is contrasted by the R-type case in Fig.~\ref{fig4}b: The PL emission in the energy range $1.43-1.46$~eV assigned to $R_h^M$ atomic registry is nearly zero at small $k$-vectors and increases with increasing $k$ as a hallmark of out-of-plane dipole moment orientation \cite{li2019direct}. With the detection angle of $54$° of our objective, we observe the PL from out-of-plane oriented interlayer excitons in $R_h^M$ atomic registry, which are dark at normal incidence and exhibit $z$-polarized emission according to dipolar selection rules \cite{forg2019cavity}. 

To complete the correspondence between the signatures of interlayer excitons in stamping-assembled \cite{zhao2022excitons} and CVD-synthesized heterostacks studied here, we discuss the results of magneto-luminescence experiments in Faraday configuration shown in Fig.~\ref{fig4}c, d for H- and R-type stacks, respectively. In finite magnetic fields, the interlayer triplet and singlet excitons in domains of $H_h^h$ atomic registry feature circularly polarized transitions, with linear valley Zeeman splitting given by $\Delta_z=E^+-E^-=g\mu_B B$ (with magnetic field $B$, Bohr magneton $\mu_B$, exciton Land\'e factor $g$, and $E^{\pm}$ referring to the dispersion branch measured under $\sigma^+$ and $\sigma^-$ detection). The triplet and singlet transitions differ characteristically in both sign and magnitude of their exciton Land\'e $g$-factors, determined from the slopes of simultaneous linear fits to the data in Fig.~\ref{fig4}c as $-17.0\pm 0.4$ and $+11.5\pm 0.3$, respectively, in agreement with DFT calculations \cite{wozniak2020exciton,forg2021moire,zhao2022excitons} and previous experiments \cite{Nagler2017,Wang2020,Delhomme2020,seyler2019signatures,brotons2020spin,joe2021electrically,zhao2022excitons}. 
In the R-type case, the magnetic field dependence of the $z$-polarized $R_h^M$ interlayer exciton is given by $E^{\pm}=E_0 \pm \frac{1}{2}\sqrt{\delta^2+\Delta_z^2}$ with zero-field exciton energy without exchange interaction $E_0$ and exchange splitting $\delta$ \cite{robert2017fine}. In Fig.~\ref{fig4}d, both dispersion branches are observed in both circular polarizations, and respecting the sign convention we thus obtain from the fit to the data with $\delta=0.1~\text{meV}$ the absolute value of the exciton Land\'e factor as $|g|=6.5$. The $g$-factor is in quantitative agreement with DFT, predicting a value of $6.3$ \cite{forg2021moire,zhao2022excitons}. In the central triangle (data not shown), we determine for $R_h^X$ an interlayer exciton $g$-factor of $+7.0 \pm 0.2$  in agreement with previous experiments and theory \cite{wozniak2020exciton,seyler2019signatures,joe2021electrically,zhao2022excitons}.

\begin{figure}[t!]
\includegraphics[scale=1.0]{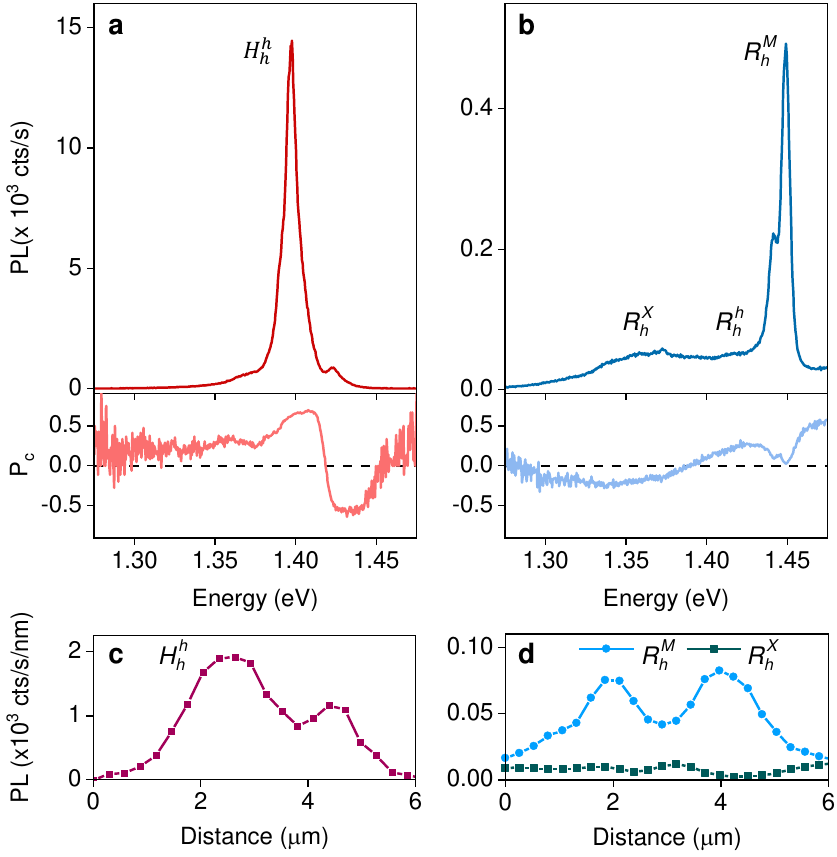}
\caption{\textbf{Spectral characteristics of grain boundaries.} \textbf{a}, \textbf{b} Top and bottom panels: Interlayer exciton PL  spectra and degree of circular polarization $P_\text{c}$ at grain boundaries in H- and R-type samples, respectively. \textbf{c}, \textbf{d} Spatial variations in the interlayer exciton PL of different atomic registries upon lateral transition from one bright spot to another indicated in the maps of Fig.~3. All measurements were performed at $100~\mu$W excitation power and $4$~K.} \label{fig5}
\end{figure}

Lastly, we note that in between extended bright areas in the PL maps of Fig.~\ref{fig3}a, b, we observed reduced PL intensities, which we ascribe to grain boundaries separating reconstructed 2D domains. Figure~\ref{fig5}a shows PL and $P_\text{c}$ spectra at such a dark position in the H-type stack with $H_h^h$ triplet and singlet exciton characteristics and slightly reduced $P_\text{c}$ as compared to a bright spot with data in Fig.~\ref{fig3}c. The variations in the PL intensity of $H_h^h$ interlayer excitons in Fig.~\ref{fig5}c is consistent with local reduction in emission (dip around $4~\mu$m with width given by the optical spot) in between two bright areas separated by a grain boundary. At the corresponding positions with reduced PL in the R-type stack, we observe additional contributions from $R_h^X$ and $R_h^h$ atomic registries with spectra in Fig.~\ref{fig5}b and negative and positive $P_\text{c}$, respectively \cite{zhao2022excitons,forg2019cavity}. The spatial intensity profiles of $R_h^M$ and $R_h^X$ interlayer exciton PL in Fig.~\ref{fig5}d are anti-correlated, identifying grain boundaries as mutually exclusive areas of competing registries with additional contribution from $R_h^h$ interlayer excitons via hot luminescence

The comparison of spectral characteristics of CVD-grown MoSe$_2$-WSe$_2$ heterostacks with our prior work on mechnically stacked samples \cite{zhao2022excitons} yields qualitatively equivalent results anticipated from general theoretical considerations. In H-type stacks, we found domains of $H_h^h$ atomic registry to dominate extended areas of heterostacks, encompassing a triangular core of 0D domains. In R-type stacks, $R_h^M$ and $R_h^X$ domains coexist in triangular cores of CVD-heteroflakes and compete for mutually exclusive reconstruction in the surrounding domains. In R-type stacks with extended domains of $R_h^M$ atomic registries, we identify the spectral signatures of interlayer excitons not reported previously from mechanically stacked samples. We anticipate that in CVD-grown heterostacks reconstruction takes place during high-temperature synthesis and thus is robust against post-processing steps of sample transfer or annealing at much lower temperatures. With extended reconstructed 2D domains and 0D arrays in their inner cores, CVD-grown MoSe$_2$-WSe$_2$ heterostacks realize areas of both moir\'e-free and moir\'e-like systems. While the latter are technologically viable as quantum dot arrays of exciton-confining potentials on the nanoscale \cite{yu2017moire,wu2018theory} and the former for dipolar exciton circuitry and extrinsic gate-modulation of potential landscapes \cite{Butov2017,Jiang2021,Ciarrocchi2022}, grain boundaries between domains of different atomic registries could host excitons with topological protection \cite{Chaves2022}.



\vspace{11pt}
\noindent \textbf{Author Information}

\noindent \textbf{Corresponding authors:}\\ F.\,T.-V. (f.tabataba@lmu.de) and A.\,H. (alexander.hoegele@lmu.de). 

\noindent \textbf{Author Contributions:}\\
Z.\,L., F.\,T.-V. and S.\,Z. contributed equally to this work.

\noindent \textbf{Notes:}\\
The authors declare no competing financial interest. 

\vspace{11pt}
\noindent \textbf{Acknowledgements:} This research was funded by the European Research Council (ERC) under the Grant Agreement No.~772195 as well as the Deutsche Forschungsgemeinschaft (DFG, German Research Foundation) within the Priority Programme SPP~2244 2DMP and the Germany's Excellence Strategy under grant No.~EXC-2111-390814868 and EXC-2089-390776260. Z.\,Li. was supported by the China Scholarship Council (CSC), grant No.~201808140196. F.\,T.-V. acknowledges funding from Munich Center for Quantum Science and Technology and the European Union's Framework Programme for Research and Innovation Horizon Europe under the Marie Sk{\l}odowska-Curie Actions grant agreement No.~101058981. S.\,Z. and I.\,B. acknowledge funding from the Alexander von Humboldt Foundation, and A.\,S.\,B. from the European Union's Framework Programme for Research and Innovation Horizon 2020 (2014--2020) under the Marie Sk{\l}odowska-Curie grant agreement No.~754388 (LMUResearchFellows) and from LMUexcellent, funded by the Federal Ministry of Education and Research (BMBF) and the Free State of Bavaria under the Excellence Strategy of the German Federal Government and the L{\"a}nder. A.\,R., K.\,M.-C. and A.\,H. acknowledge funding by the Bavarian Hightech Agenda within the Munich Quantum Valley doctoral fellowship program and the EQAP project. G.\,R.\,S. acknowledges funding from the Army Research Office under Cooperative Agreement No. W911NF-21-2-0147. E.\,K. acknowledges funding from the STC Center for Integrated Quantum Materials, NSF Grant No. DMR-1231319, NSF DMREF Award No. 1922172, and the Army Research Office under Cooperative Agreement No. W911NF-21-2-0147. K.\,W. and T.\,T. acknowledge support from JSPS KAKENHI (grant No. 19H05790, 20H00354 and 21H05233).

\pagebreak
\newpage
\widetext
\begin{center}
\textbf{\Large Lattice reconstruction in MoSe$_2$-WSe$_2$ heterobilayers
synthesized by chemical vapor deposition\\ \vspace{6pt} \Large{Supporting Information}}
\end{center}

\noindent 
{\textbf{\large{Supporting Note 1: Synthesis of MoSe$_2$-WSe$_2$ heterobilayers}}

\noindent MoSe$_2$-WSe$_2$ heterobilayers were grown by a modified two-step CVD method on thermally oxidized silicon substrates (with a SiO$_2$ thickness of $285$~nm). A three-zone furnace CVD system (Carbolite Gero) equipped with a 1-inch quartz tube was used for the growth. First, MoSe$_2$ monolayers were grown employing MoO$_2$ and Se powders (99.99\%, Sigma-Aldrich) as precursors and the vapor phase chalcogenization method to obtain high-quality crystalline samples \cite{bilgin2015chemical}. An aluminum boat with MoO$_2$ powder was placed at the center of the first heating zone and a SiO$_2$/Si substrate was suspended face-down above it. Another crucible with Se powder was placed $10$~cm upstream from the center. After the tube was evacuated to $10$~mTorr several times to remove air and moisture, the reaction chamber pressure was increased to ambient pressure through $300$~sccm of ultra-high purity Ar flowed into the chamber to act as a carrier gas and to create an inert atmosphere. The furnace temperature was raised to $750$°C at a rate of $10$°C/min and maintained for $15$~min for the growth. $0.75$~sccm of H$_2$ was introduced as a reductant gas when the temperature reached $750$°C. The system was cooled down to room temperature after the growth. The as-grown MoSe$_2$ monolayers were used as the substrate for the vertical growth of the WSe$_2$ monolayers. The WSe$_2$ powder (99.8\%, Alfa Aesar) and the substrate were placed at the center and downstream of the quartz tube, respectively. While increasing the temperature, a reverse flow of $100$~sccm cold Ar gas cooled the MoSe$_2$ monolayers to counteract thermal degradation. The center zone of the furnace was heated to $1090$°C under atmospheric pressure. After reaching the desired temperature for WSe$_2$ growth, the inverse Ar gas flow was switched to a mixture of Ar/H$_2$ (with 2\% H$_2$) forward gas flow to carry the vaporized WSe$_2$ precursor to the substrate. Then, the growth temperature was maintained for $3$~min to synthesize WSe$_2$. After the growth, the furnace was cooled down to room temperature.

\vspace{11pt}
\noindent 
{\textbf{\large{Supporting Note 2: Sample fabrication}}

\noindent A PDMS/PC stamp was used to sequentially pick up the exfoliated hBN layers (NIMS) and CVD-grown  MoSe$_2$-WSe$_2$ heterobilayer using the dry transfer method \cite{pizzocchero2016hot,purdie2018cleaning}. Poly-(Bisphenol A-carbonate) pellets (Sigma Aldrich) were dissolved in chloroform with a weight ratio of 8. The mixture was stirred at 500 rpm using a magneton bar at room temperature overnight. The well-dissolved PC film was mounted on a PDMS block on a glass slide. First, the top hBN layer with a thickness of 140 nm was picked up with the stamp, followed by the heterobilayer and the bottom hBN layer with a thickness of 100 nm. The pick-up temperatures for the hBN flakes and heterobilayer were around 50°C and 140°C, respectively. The entire stack was released at a temperature of 170°C onto a SiO$_2$/Si target substrate, then soaked in chloroform solution for 20 min to remove PC residues, cleaned by acetone and isopropanol and annealed at 200°C under ultrahigh vacuum for 15 hours.

\vspace{11pt}
\noindent 
{\textbf{\large{Supporting Note 3: SEM imaging}}

\noindent SEM imaging of MoSe$_2$-WSe$_2$ heterobilayers was performed with a Raith eLine system equipped with an Everhart-Thornley detector using the secondary electron imaging technique \cite{ashida2015crystallographic,andersen2021excitons}. In a heterobilayer, the two hexagonal lattices form an effective cavity for electrons, resulting in an atomic registry dependent generation of secondary electrons. Under optimized experimental  conditions, the different atomic registries can be differentiated by the contrast in secondary electron imaging. The measurements were performed at 1 keV electron beam energy and normal incidence for H-type and 38° tilt for R-type samples \cite{zhao2022excitons}.

\vspace{11pt}
\noindent 
{\textbf{\large{Supporting Note 4: Theoretical modeling}}

\noindent Mechanical properties were obtained from the generalized stacking fault energy (GSFE) \cite{carr2018relaxation}, extracting its parameters from DFT calculations as implemented in the VASP \cite{Kresse1996} code. We used the r$^2$SCAN+rVV10 exchange-correlation functional \cite{Furness2020}, a meta-GGA functional that accurately describes structural and electronic properties accounting for van der Waals interactions \cite{Ning2022}. All calculations used projector-augmented wave (PAW) pseudopotentials \cite{Bloechl1994}, plane-wave expansion of the wavefunctions with an energy cutoff of $292$~eV, \textbf{k}-point sampling by  $21 \times 21~\Gamma$-centered Monkhorst-Pack meshes, and electronic self-consistent convergence of $10^{-6}$~eV. Atomic positions were relaxed until Hellmann-Feynman forces were smaller than $10^{-2}$~eV $\AA^{-1}$. We add a vacuum layer of at least 15 $\AA$ in the z-direction, and include dipole corrections. First, we obtained the optimized MoSe$_2$-WSe$_2$ heterostructure lattice parameters as $3.289$ and $3.287 \AA$ for the R- and H-stacking, respectively, and then we calculated total energies including vertical relaxation for a grid of $9$ displaced unit cells, where one of the layers was displaced along the in-plane diagonal relative to the other. For the R-stacking, the calculated interlayer distance between the metal atoms was $6.56$ and 7.15 $\AA$ at the minimum and maximum energy stackings, respectively, and $6.55$ and $7.063 \AA$ for the H-stacking. The resulting fitted GSFE is shown in Fig.~1d, e.

\vspace{11pt}
\noindent 
{\textbf{\large{Supporting Note 5: STEM imaging}}

\noindent CVD-grown heterobilayers were transferred to Quantifoil® grids and analyzed with a probe corrected FEI Titan Themis scanning transmission electron microscope (STEM) at an acceleration voltage of $300$~keV equipped with a monochromator for the illumination. High-resolution high-angle annular dark-field (HAADF) imaging was conducted using a Fischione Model 3000 HAADF detector with a camera length of $195$~mm and a convergence semi-angle of $23$~mrad. To allow direct comparison with the experiment, multislice simulations of HAADF-STEM images were carried out using the experimental parameters and 50 frozen phonon configurations, and one slice per atomic layer. The phase gratings for the simulations were created by stacking of monolayers of MoSe$_2$ and WSe$_2$ followed by a relative rotation and shift in order to generate expected heterobilayer stacking configurations. For each heterobilayer region, an atomically-resolved image of the adjacent MoSe$_2$ monolayer was recorded in order to verify the relative crystallographic orientation of the two regions. As care must be taken to not cause significant specimen damage at $300$~keV beam energy, the intensity of the electron beam was reduced by defocusing the monochromator slightly. To improve the signal-to-noise ratio, the images have been segmented into translational invariant cells based on the atomic peaks from the Z-contrast data. Within each cell, a geometric transform was determined to account for scanning distortions and drift gradients, and an average cell was determined as shown in the insets in Fig.~2 of the main text.

\vspace{11pt}
\noindent 
{\textbf{\large{Supporting Note 6: Optical spectroscopy}}

\noindent Cryogenic PL and DR measurements were performed in back-scattering geometry in a close-cycle cryostat (attocube systems, attoDRY800 or attoDRY1000) with a base temperature of $\sim 4$~K and a solenoid with magnetic fields of up to $\pm9$~T in Faraday configuration. Sample positioning was performed with respect to a low-temperature apochromatic objective (attocube systems, LT-APO/NIR/0.81 or 633-RAMAN/0.81) with piezo-units (attocube systems, ANPx101, ANPz101, and ANSxy100). PL was excited at $725$~nm with a Ti:sapphire laser (Coherent, Mira or SolsTiS, M Squared) in continuous-wave mode. DR, defined as $DR = (R-R_0)/R_0$, where $R$ was the reflectance from the sample and $R_0$ was the reference reflectance on the nearby substrate with hBN, was recorded with a Tungsten-Halogen lamp (Thorlabs, SLS201L or Ocean Insight, HL-2000-HP). The signal was dispersed by a monochromator (Roper Scientific, Acton SpectraPro 300i or Teledyne Princton Instruments, IsoPlane SCT320) with a $300$ or $1200$~grooves/mm grating and detected by a Peltier-cooled CCD (Andor, iDus 416 or Teledyne Princton Instruments, PIXIS 1024). A set of linear polarizers (Thorlabs, LPVIS), half- and quarter-waveplates (B. Halle, $310-1100$~nm achromatic) mounted on piezo-rotators (attocube systems, ANR240) were used to control the polarization in excitation and detection. The degree of circular polarization, defined as $P_\text{c} = (I_{\text{co}}-I_{\text{cross}})/(I_{\text{co}}+I_{\text{cross}})$, was determined by measuring $I_{\text{co}}$ and $I_{\text{cross}}$ under circularly polarized excitation and co- and cross-polarized detection. Momentum-space imaging in 4f and telescope configuration employed four achromatic doublet lenses (Edmund Optics, VIS-NIR).


\begin{thebibliography}{47}%
\makeatletter
\providecommand \@ifxundefined [1]{%
 \@ifx{#1\undefined}
}%
\providecommand \@ifnum [1]{%
 \ifnum #1\expandafter \@firstoftwo
 \else \expandafter \@secondoftwo
 \fi
}%
\providecommand \@ifx [1]{%
 \ifx #1\expandafter \@firstoftwo
 \else \expandafter \@secondoftwo
 \fi
}%
\providecommand \natexlab [1]{#1}%
\providecommand \enquote  [1]{``#1''}%
\providecommand \bibnamefont  [1]{#1}%
\providecommand \bibfnamefont [1]{#1}%
\providecommand \citenamefont [1]{#1}%
\providecommand \href@noop [0]{\@secondoftwo}%
\providecommand \href [0]{\begingroup \@sanitize@url \@href}%
\providecommand \@href[1]{\@@startlink{#1}\@@href}%
\providecommand \@@href[1]{\endgroup#1\@@endlink}%
\providecommand \@sanitize@url [0]{\catcode `\\12\catcode `\$12\catcode
  `\&12\catcode `\#12\catcode `\^12\catcode `\_12\catcode `\%12\relax}%
\providecommand \@@startlink[1]{}%
\providecommand \@@endlink[0]{}%
\providecommand \url  [0]{\begingroup\@sanitize@url \@url }%
\providecommand \@url [1]{\endgroup\@href {#1}{\urlprefix }}%
\providecommand \urlprefix  [0]{URL }%
\providecommand \Eprint [0]{\href }%
\providecommand \doibase [0]{https://doi.org/}%
\providecommand \selectlanguage [0]{\@gobble}%
\providecommand \bibinfo  [0]{\@secondoftwo}%
\providecommand \bibfield  [0]{\@secondoftwo}%
\providecommand \translation [1]{[#1]}%
\providecommand \BibitemOpen [0]{}%
\providecommand \bibitemStop [0]{}%
\providecommand \bibitemNoStop [0]{.\EOS\space}%
\providecommand \EOS [0]{\spacefactor3000\relax}%
\providecommand \BibitemShut  [1]{\csname bibitem#1\endcsname}%
\let\auto@bib@innerbib\@empty
\bibitem [{\citenamefont {Wu}\ \emph {et~al.}(2018{\natexlab{a}})\citenamefont
  {Wu}, \citenamefont {Lovorn}, \citenamefont {Tutuc},\ and\ \citenamefont
  {MacDonald}}]{wu2018hubbard}%
  \BibitemOpen
  \bibfield  {author} {\bibinfo {author} {\bibfnamefont {F.}~\bibnamefont
  {Wu}}, \bibinfo {author} {\bibfnamefont {T.}~\bibnamefont {Lovorn}}, \bibinfo
  {author} {\bibfnamefont {E.}~\bibnamefont {Tutuc}},\ and\ \bibinfo {author}
  {\bibfnamefont {A.~H.}\ \bibnamefont {MacDonald}},\ }\bibfield  {title}
  {\bibinfo {title} {Hubbard model physics in transition metal dichalcogenide
  moir{\'e} bands},\ }\href@noop {} {\bibfield  {journal} {\bibinfo  {journal}
  {Phys. Rev. Lett.}\ }\textbf {\bibinfo {volume} {121}},\ \bibinfo {pages}
  {026402} (\bibinfo {year} {2018}{\natexlab{a}})}\BibitemShut {NoStop}%
\bibitem [{\citenamefont {Tang}\ \emph {et~al.}(2020)\citenamefont {Tang},
  \citenamefont {Li}, \citenamefont {Li}, \citenamefont {Xu}, \citenamefont
  {Liu}, \citenamefont {Barmak}, \citenamefont {Watanabe}, \citenamefont
  {Taniguchi}, \citenamefont {MacDonald}, \citenamefont {Shan},\ and\
  \citenamefont {Mak}}]{tang2020simulation}%
  \BibitemOpen
  \bibfield  {author} {\bibinfo {author} {\bibfnamefont {Y.}~\bibnamefont
  {Tang}}, \bibinfo {author} {\bibfnamefont {L.}~\bibnamefont {Li}}, \bibinfo
  {author} {\bibfnamefont {T.}~\bibnamefont {Li}}, \bibinfo {author}
  {\bibfnamefont {Y.}~\bibnamefont {Xu}}, \bibinfo {author} {\bibfnamefont
  {S.}~\bibnamefont {Liu}}, \bibinfo {author} {\bibfnamefont {K.}~\bibnamefont
  {Barmak}}, \bibinfo {author} {\bibfnamefont {K.}~\bibnamefont {Watanabe}},
  \bibinfo {author} {\bibfnamefont {T.}~\bibnamefont {Taniguchi}}, \bibinfo
  {author} {\bibfnamefont {A.~H.}\ \bibnamefont {MacDonald}}, \bibinfo {author}
  {\bibfnamefont {J.}~\bibnamefont {Shan}},\ and\ \bibinfo {author}
  {\bibfnamefont {K.~F.}\ \bibnamefont {Mak}},\ }\bibfield  {title} {\bibinfo
  {title} {{Simulation of Hubbard model physics in {WSe}$_2$/{WS}$_2$ moir{\'e}
  superlattices}},\ }\href@noop {} {\bibfield  {journal} {\bibinfo  {journal}
  {Nature}\ }\textbf {\bibinfo {volume} {579}},\ \bibinfo {pages} {353}
  (\bibinfo {year} {2020})}\BibitemShut {NoStop}%
\bibitem [{\citenamefont {Shimazaki}\ \emph {et~al.}(2020)\citenamefont
  {Shimazaki}, \citenamefont {Schwartz}, \citenamefont {Watanabe},
  \citenamefont {Taniguchi}, \citenamefont {Kroner},\ and\ \citenamefont
  {Imamo{\u{g}}lu}}]{shimazaki2020strongly}%
  \BibitemOpen
  \bibfield  {author} {\bibinfo {author} {\bibfnamefont {Y.}~\bibnamefont
  {Shimazaki}}, \bibinfo {author} {\bibfnamefont {I.}~\bibnamefont {Schwartz}},
  \bibinfo {author} {\bibfnamefont {K.}~\bibnamefont {Watanabe}}, \bibinfo
  {author} {\bibfnamefont {T.}~\bibnamefont {Taniguchi}}, \bibinfo {author}
  {\bibfnamefont {M.}~\bibnamefont {Kroner}},\ and\ \bibinfo {author}
  {\bibfnamefont {A.}~\bibnamefont {Imamo{\u{g}}lu}},\ }\bibfield  {title}
  {\bibinfo {title} {Strongly correlated electrons and hybrid excitons in a
  moir{\'e} heterostructure},\ }\href@noop {} {\bibfield  {journal} {\bibinfo
  {journal} {Nature}\ }\textbf {\bibinfo {volume} {580}},\ \bibinfo {pages}
  {472} (\bibinfo {year} {2020})}\BibitemShut {NoStop}%
\bibitem [{\citenamefont {Regan}\ \emph {et~al.}(2020)\citenamefont {Regan},
  \citenamefont {Wang}, \citenamefont {Jin}, \citenamefont {Bakti~Utama},
  \citenamefont {Gao}, \citenamefont {Wei}, \citenamefont {Zhao}, \citenamefont
  {Zhao}, \citenamefont {Zhang}, \citenamefont {Yumigeta}, \citenamefont
  {Blei}, \citenamefont {Carlström}, \citenamefont {Watanabe}, \citenamefont
  {Taniguchi}, \citenamefont {Tongay}, \citenamefont {Crommie}, \citenamefont
  {Zettl},\ and\ \citenamefont {Wang}}]{regan2020mott}%
  \BibitemOpen
  \bibfield  {author} {\bibinfo {author} {\bibfnamefont {E.~C.}\ \bibnamefont
  {Regan}}, \bibinfo {author} {\bibfnamefont {D.}~\bibnamefont {Wang}},
  \bibinfo {author} {\bibfnamefont {C.}~\bibnamefont {Jin}}, \bibinfo {author}
  {\bibfnamefont {M.~I.}\ \bibnamefont {Bakti~Utama}}, \bibinfo {author}
  {\bibfnamefont {B.}~\bibnamefont {Gao}}, \bibinfo {author} {\bibfnamefont
  {X.}~\bibnamefont {Wei}}, \bibinfo {author} {\bibfnamefont {S.}~\bibnamefont
  {Zhao}}, \bibinfo {author} {\bibfnamefont {W.}~\bibnamefont {Zhao}}, \bibinfo
  {author} {\bibfnamefont {Z.}~\bibnamefont {Zhang}}, \bibinfo {author}
  {\bibfnamefont {K.}~\bibnamefont {Yumigeta}}, \bibinfo {author}
  {\bibfnamefont {M.}~\bibnamefont {Blei}}, \bibinfo {author} {\bibfnamefont
  {J.~D.}\ \bibnamefont {Carlström}}, \bibinfo {author} {\bibfnamefont
  {K.}~\bibnamefont {Watanabe}}, \bibinfo {author} {\bibfnamefont
  {T.}~\bibnamefont {Taniguchi}}, \bibinfo {author} {\bibfnamefont
  {S.}~\bibnamefont {Tongay}}, \bibinfo {author} {\bibfnamefont
  {M.}~\bibnamefont {Crommie}}, \bibinfo {author} {\bibfnamefont
  {A.}~\bibnamefont {Zettl}},\ and\ \bibinfo {author} {\bibfnamefont
  {F.}~\bibnamefont {Wang}},\ }\bibfield  {title} {\bibinfo {title} {Mott and
  generalized {W}igner crystal states in {WSe}$_2$/{WS}$_2$ moir{\'e}
  superlattices},\ }\href@noop {} {\bibfield  {journal} {\bibinfo  {journal}
  {Nature}\ }\textbf {\bibinfo {volume} {579}},\ \bibinfo {pages} {359}
  (\bibinfo {year} {2020})}\BibitemShut {NoStop}%
\bibitem [{\citenamefont {Xu}\ \emph {et~al.}(2020)\citenamefont {Xu},
  \citenamefont {Liu}, \citenamefont {Rhodes}, \citenamefont {Watanabe},
  \citenamefont {Taniguchi}, \citenamefont {Hone}, \citenamefont {Elser},
  \citenamefont {Mak},\ and\ \citenamefont {Shan}}]{xu2020correlated}%
  \BibitemOpen
  \bibfield  {author} {\bibinfo {author} {\bibfnamefont {Y.}~\bibnamefont
  {Xu}}, \bibinfo {author} {\bibfnamefont {S.}~\bibnamefont {Liu}}, \bibinfo
  {author} {\bibfnamefont {D.~A.}\ \bibnamefont {Rhodes}}, \bibinfo {author}
  {\bibfnamefont {K.}~\bibnamefont {Watanabe}}, \bibinfo {author}
  {\bibfnamefont {T.}~\bibnamefont {Taniguchi}}, \bibinfo {author}
  {\bibfnamefont {J.}~\bibnamefont {Hone}}, \bibinfo {author} {\bibfnamefont
  {V.}~\bibnamefont {Elser}}, \bibinfo {author} {\bibfnamefont {K.~F.}\
  \bibnamefont {Mak}},\ and\ \bibinfo {author} {\bibfnamefont {J.}~\bibnamefont
  {Shan}},\ }\bibfield  {title} {\bibinfo {title} {Correlated insulating states
  at fractional fillings of moir{\'e} superlattices},\ }\href@noop {}
  {\bibfield  {journal} {\bibinfo  {journal} {Nature}\ }\textbf {\bibinfo
  {volume} {587}},\ \bibinfo {pages} {214} (\bibinfo {year}
  {2020})}\BibitemShut {NoStop}%
\bibitem [{\citenamefont {Tang}\ \emph {et~al.}(2022)\citenamefont {Tang},
  \citenamefont {Gu}, \citenamefont {Liu}, \citenamefont {Watanabe},
  \citenamefont {Taniguchi}, \citenamefont {Hone}, \citenamefont {Mak},\ and\
  \citenamefont {Shan}}]{tang2022dielectric}%
  \BibitemOpen
  \bibfield  {author} {\bibinfo {author} {\bibfnamefont {Y.}~\bibnamefont
  {Tang}}, \bibinfo {author} {\bibfnamefont {J.}~\bibnamefont {Gu}}, \bibinfo
  {author} {\bibfnamefont {S.}~\bibnamefont {Liu}}, \bibinfo {author}
  {\bibfnamefont {K.}~\bibnamefont {Watanabe}}, \bibinfo {author}
  {\bibfnamefont {T.}~\bibnamefont {Taniguchi}}, \bibinfo {author}
  {\bibfnamefont {J.~C.}\ \bibnamefont {Hone}}, \bibinfo {author}
  {\bibfnamefont {K.~F.}\ \bibnamefont {Mak}},\ and\ \bibinfo {author}
  {\bibfnamefont {J.}~\bibnamefont {Shan}},\ }\bibfield  {title} {\bibinfo
  {title} {Dielectric catastrophe at the {M}ott and {W}igner transitions in a
  moiré superlattice},\ }\href@noop {} {\bibfield  {journal} {\bibinfo
  {journal} {Nat. Commun.}\ }\textbf {\bibinfo {volume} {13}},\ \bibinfo
  {pages} {4271} (\bibinfo {year} {2022})}\BibitemShut {NoStop}%
\bibitem [{\citenamefont {Zhang}\ \emph {et~al.}(2018)\citenamefont {Zhang},
  \citenamefont {Surrente}, \citenamefont {Baranowski}, \citenamefont {Maude},
  \citenamefont {Gant}, \citenamefont {Castellanos-Gomez},\ and\ \citenamefont
  {Plochocka}}]{zhang2018moire}%
  \BibitemOpen
  \bibfield  {author} {\bibinfo {author} {\bibfnamefont {N.}~\bibnamefont
  {Zhang}}, \bibinfo {author} {\bibfnamefont {A.}~\bibnamefont {Surrente}},
  \bibinfo {author} {\bibfnamefont {M.}~\bibnamefont {Baranowski}}, \bibinfo
  {author} {\bibfnamefont {D.~K.}\ \bibnamefont {Maude}}, \bibinfo {author}
  {\bibfnamefont {P.}~\bibnamefont {Gant}}, \bibinfo {author} {\bibfnamefont
  {A.}~\bibnamefont {Castellanos-Gomez}},\ and\ \bibinfo {author}
  {\bibfnamefont {P.}~\bibnamefont {Plochocka}},\ }\bibfield  {title} {\bibinfo
  {title} {Moir{\'e} intralayer excitons in a {MoSe}$_2$/{MoS}$_2$
  heterostructure},\ }\href@noop {} {\bibfield  {journal} {\bibinfo  {journal}
  {Nano Lett.}\ }\textbf {\bibinfo {volume} {18}},\ \bibinfo {pages} {7651}
  (\bibinfo {year} {2018})}\BibitemShut {NoStop}%
\bibitem [{\citenamefont {Naik}\ \emph {et~al.}(2022)\citenamefont {Naik},
  \citenamefont {Regan}, \citenamefont {Zhang}, \citenamefont {Chan},
  \citenamefont {Li}, \citenamefont {Wang}, \citenamefont {Yoon}, \citenamefont
  {Ong}, \citenamefont {Zhao}, \citenamefont {Zhao}, \citenamefont {Utam},
  \citenamefont {Gao}, \citenamefont {Wei}, \citenamefont {Sayyad},
  \citenamefont {Yumigeta}, \citenamefont {Watanabe}, \citenamefont
  {Taniguchi}, \citenamefont {Tongay}, \citenamefont {da~Jornada},
  \citenamefont {Wang},\ and\ \citenamefont {Louie}}]{naik2022intralayer}%
  \BibitemOpen
  \bibfield  {author} {\bibinfo {author} {\bibfnamefont {M.~H.}\ \bibnamefont
  {Naik}}, \bibinfo {author} {\bibfnamefont {E.~C.}\ \bibnamefont {Regan}},
  \bibinfo {author} {\bibfnamefont {Z.}~\bibnamefont {Zhang}}, \bibinfo
  {author} {\bibfnamefont {Y.-H.}\ \bibnamefont {Chan}}, \bibinfo {author}
  {\bibfnamefont {Z.}~\bibnamefont {Li}}, \bibinfo {author} {\bibfnamefont
  {D.}~\bibnamefont {Wang}}, \bibinfo {author} {\bibfnamefont {Y.}~\bibnamefont
  {Yoon}}, \bibinfo {author} {\bibfnamefont {C.~S.}\ \bibnamefont {Ong}},
  \bibinfo {author} {\bibfnamefont {W.}~\bibnamefont {Zhao}}, \bibinfo {author}
  {\bibfnamefont {S.}~\bibnamefont {Zhao}}, \bibinfo {author} {\bibfnamefont
  {M.~I.~B.}\ \bibnamefont {Utam}}, \bibinfo {author} {\bibfnamefont
  {B.}~\bibnamefont {Gao}}, \bibinfo {author} {\bibfnamefont {X.}~\bibnamefont
  {Wei}}, \bibinfo {author} {\bibfnamefont {M.}~\bibnamefont {Sayyad}},
  \bibinfo {author} {\bibfnamefont {K.}~\bibnamefont {Yumigeta}}, \bibinfo
  {author} {\bibfnamefont {K.}~\bibnamefont {Watanabe}}, \bibinfo {author}
  {\bibfnamefont {T.}~\bibnamefont {Taniguchi}}, \bibinfo {author}
  {\bibfnamefont {S.}~\bibnamefont {Tongay}}, \bibinfo {author} {\bibfnamefont
  {F.~H.}\ \bibnamefont {da~Jornada}}, \bibinfo {author} {\bibfnamefont
  {F.}~\bibnamefont {Wang}},\ and\ \bibinfo {author} {\bibfnamefont {S.~G.}\
  \bibnamefont {Louie}},\ }\bibfield  {title} {\bibinfo {title} {{Intralayer
  charge-transfer moir{\'e} excitons in van der Waals superlattices}},\
  }\href@noop {} {\bibfield  {journal} {\bibinfo  {journal} {Nature}\ }\textbf
  {\bibinfo {volume} {609}},\ \bibinfo {pages} {52} (\bibinfo {year}
  {2022})}\BibitemShut {NoStop}%
\bibitem [{\citenamefont {Seyler}\ \emph {et~al.}(2019)\citenamefont {Seyler},
  \citenamefont {Rivera}, \citenamefont {Yu}, \citenamefont {Wilson},
  \citenamefont {Ray}, \citenamefont {Mandrus}, \citenamefont {Yan},
  \citenamefont {Yao},\ and\ \citenamefont {Xu}}]{seyler2019signatures}%
  \BibitemOpen
  \bibfield  {author} {\bibinfo {author} {\bibfnamefont {K.~L.}\ \bibnamefont
  {Seyler}}, \bibinfo {author} {\bibfnamefont {P.}~\bibnamefont {Rivera}},
  \bibinfo {author} {\bibfnamefont {H.}~\bibnamefont {Yu}}, \bibinfo {author}
  {\bibfnamefont {N.~P.}\ \bibnamefont {Wilson}}, \bibinfo {author}
  {\bibfnamefont {E.~L.}\ \bibnamefont {Ray}}, \bibinfo {author} {\bibfnamefont
  {D.~G.}\ \bibnamefont {Mandrus}}, \bibinfo {author} {\bibfnamefont
  {J.}~\bibnamefont {Yan}}, \bibinfo {author} {\bibfnamefont {W.}~\bibnamefont
  {Yao}},\ and\ \bibinfo {author} {\bibfnamefont {X.}~\bibnamefont {Xu}},\
  }\bibfield  {title} {\bibinfo {title} {Signatures of moir{\'e}-trapped valley
  excitons in {MoSe}$_2$/{WSe}$_2$ heterobilayers},\ }\href@noop {} {\bibfield
  {journal} {\bibinfo  {journal} {Nature}\ }\textbf {\bibinfo {volume} {567}},\
  \bibinfo {pages} {66} (\bibinfo {year} {2019})}\BibitemShut {NoStop}%
\bibitem [{\citenamefont {Tran}\ \emph {et~al.}(2019)\citenamefont {Tran},
  \citenamefont {Moody}, \citenamefont {Wu}, \citenamefont {Lu}, \citenamefont
  {Choi}, \citenamefont {Kim}, \citenamefont {Rai}, \citenamefont {Sanchez},
  \citenamefont {Quan}, \citenamefont {Singh}, \citenamefont {Embley},
  \citenamefont {Zepeda}, \citenamefont {Campbell}, \citenamefont {Autry},
  \citenamefont {Taniguchi}, \citenamefont {Watanabe}, \citenamefont {Lu},
  \citenamefont {Banerjee}, \citenamefont {Silverman}, \citenamefont {Kim},
  \citenamefont {Tutuc}, \citenamefont {Yang}, \citenamefont {MacDonald},\ and\
  \citenamefont {Li}}]{tran2019evidence}%
  \BibitemOpen
  \bibfield  {author} {\bibinfo {author} {\bibfnamefont {K.}~\bibnamefont
  {Tran}}, \bibinfo {author} {\bibfnamefont {G.}~\bibnamefont {Moody}},
  \bibinfo {author} {\bibfnamefont {F.}~\bibnamefont {Wu}}, \bibinfo {author}
  {\bibfnamefont {X.}~\bibnamefont {Lu}}, \bibinfo {author} {\bibfnamefont
  {J.}~\bibnamefont {Choi}}, \bibinfo {author} {\bibfnamefont {K.}~\bibnamefont
  {Kim}}, \bibinfo {author} {\bibfnamefont {A.}~\bibnamefont {Rai}}, \bibinfo
  {author} {\bibfnamefont {D.~A.}\ \bibnamefont {Sanchez}}, \bibinfo {author}
  {\bibfnamefont {J.}~\bibnamefont {Quan}}, \bibinfo {author} {\bibfnamefont
  {A.}~\bibnamefont {Singh}}, \bibinfo {author} {\bibfnamefont
  {J.}~\bibnamefont {Embley}}, \bibinfo {author} {\bibfnamefont
  {A.}~\bibnamefont {Zepeda}}, \bibinfo {author} {\bibfnamefont
  {M.}~\bibnamefont {Campbell}}, \bibinfo {author} {\bibfnamefont
  {T.}~\bibnamefont {Autry}}, \bibinfo {author} {\bibfnamefont
  {T.}~\bibnamefont {Taniguchi}}, \bibinfo {author} {\bibfnamefont
  {K.}~\bibnamefont {Watanabe}}, \bibinfo {author} {\bibfnamefont
  {N.}~\bibnamefont {Lu}}, \bibinfo {author} {\bibfnamefont {S.~K.}\
  \bibnamefont {Banerjee}}, \bibinfo {author} {\bibfnamefont {K.~L.}\
  \bibnamefont {Silverman}}, \bibinfo {author} {\bibfnamefont {S.}~\bibnamefont
  {Kim}}, \bibinfo {author} {\bibfnamefont {E.}~\bibnamefont {Tutuc}}, \bibinfo
  {author} {\bibfnamefont {L.}~\bibnamefont {Yang}}, \bibinfo {author}
  {\bibfnamefont {A.~H.}\ \bibnamefont {MacDonald}},\ and\ \bibinfo {author}
  {\bibfnamefont {X.}~\bibnamefont {Li}},\ }\bibfield  {title} {\bibinfo
  {title} {Evidence for moir{\'e} excitons in van der {W}aals
  heterostructures},\ }\href@noop {} {\bibfield  {journal} {\bibinfo  {journal}
  {Nature}\ }\textbf {\bibinfo {volume} {567}},\ \bibinfo {pages} {71}
  (\bibinfo {year} {2019})}\BibitemShut {NoStop}%
\bibitem [{\citenamefont {Jin}\ \emph {et~al.}(2019)\citenamefont {Jin},
  \citenamefont {Regan}, \citenamefont {Yan}, \citenamefont {Iqbal
  Bakti~Utama}, \citenamefont {Wang}, \citenamefont {Zhao}, \citenamefont
  {Qin}, \citenamefont {Yang}, \citenamefont {Zheng}, \citenamefont {Shi},
  \citenamefont {Watanabe}, \citenamefont {Taniguchi}, \citenamefont {Tongay},
  \citenamefont {Zettl},\ and\ \citenamefont {Wang}}]{jin2019observation}%
  \BibitemOpen
  \bibfield  {author} {\bibinfo {author} {\bibfnamefont {C.}~\bibnamefont
  {Jin}}, \bibinfo {author} {\bibfnamefont {E.~C.}\ \bibnamefont {Regan}},
  \bibinfo {author} {\bibfnamefont {A.}~\bibnamefont {Yan}}, \bibinfo {author}
  {\bibfnamefont {M.}~\bibnamefont {Iqbal Bakti~Utama}}, \bibinfo {author}
  {\bibfnamefont {D.}~\bibnamefont {Wang}}, \bibinfo {author} {\bibfnamefont
  {S.}~\bibnamefont {Zhao}}, \bibinfo {author} {\bibfnamefont {Y.}~\bibnamefont
  {Qin}}, \bibinfo {author} {\bibfnamefont {S.}~\bibnamefont {Yang}}, \bibinfo
  {author} {\bibfnamefont {Z.}~\bibnamefont {Zheng}}, \bibinfo {author}
  {\bibfnamefont {S.}~\bibnamefont {Shi}}, \bibinfo {author} {\bibfnamefont
  {K.}~\bibnamefont {Watanabe}}, \bibinfo {author} {\bibfnamefont
  {T.}~\bibnamefont {Taniguchi}}, \bibinfo {author} {\bibfnamefont
  {S.}~\bibnamefont {Tongay}}, \bibinfo {author} {\bibfnamefont
  {A.}~\bibnamefont {Zettl}},\ and\ \bibinfo {author} {\bibfnamefont
  {F.}~\bibnamefont {Wang}},\ }\bibfield  {title} {\bibinfo {title}
  {Observation of moir{\'e} excitons in {WSe}$_2$/{WS}$_2$ heterostructure
  superlattices},\ }\href@noop {} {\bibfield  {journal} {\bibinfo  {journal}
  {Nature}\ }\textbf {\bibinfo {volume} {567}},\ \bibinfo {pages} {76}
  (\bibinfo {year} {2019})}\BibitemShut {NoStop}%
\bibitem [{\citenamefont {Alexeev}\ \emph {et~al.}(2019)\citenamefont
  {Alexeev}, \citenamefont {Ruiz-Tijerina}, \citenamefont {Danovich},
  \citenamefont {Hamer}, \citenamefont {Terry}, \citenamefont {Nayak},
  \citenamefont {Ahn}, \citenamefont {Pak}, \citenamefont {Lee}, \citenamefont
  {Sohn}, \citenamefont {Molas}, \citenamefont {Koperski}, \citenamefont
  {Watanabe}, \citenamefont {Taniguchi}, \citenamefont {Novoselov},
  \citenamefont {Gorbachev}, \citenamefont {Shin}, \citenamefont {Fal’ko},\
  and\ \citenamefont {Tartakovskii}}]{alexeev2019resonantly}%
  \BibitemOpen
  \bibfield  {author} {\bibinfo {author} {\bibfnamefont {E.~M.}\ \bibnamefont
  {Alexeev}}, \bibinfo {author} {\bibfnamefont {D.~A.}\ \bibnamefont
  {Ruiz-Tijerina}}, \bibinfo {author} {\bibfnamefont {M.}~\bibnamefont
  {Danovich}}, \bibinfo {author} {\bibfnamefont {M.~J.}\ \bibnamefont {Hamer}},
  \bibinfo {author} {\bibfnamefont {D.~J.}\ \bibnamefont {Terry}}, \bibinfo
  {author} {\bibfnamefont {P.~K.}\ \bibnamefont {Nayak}}, \bibinfo {author}
  {\bibfnamefont {S.}~\bibnamefont {Ahn}}, \bibinfo {author} {\bibfnamefont
  {S.}~\bibnamefont {Pak}}, \bibinfo {author} {\bibfnamefont {J.}~\bibnamefont
  {Lee}}, \bibinfo {author} {\bibfnamefont {J.~I.}\ \bibnamefont {Sohn}},
  \bibinfo {author} {\bibfnamefont {M.~R.}\ \bibnamefont {Molas}}, \bibinfo
  {author} {\bibfnamefont {M.}~\bibnamefont {Koperski}}, \bibinfo {author}
  {\bibfnamefont {K.}~\bibnamefont {Watanabe}}, \bibinfo {author}
  {\bibfnamefont {T.}~\bibnamefont {Taniguchi}}, \bibinfo {author}
  {\bibfnamefont {K.~S.}\ \bibnamefont {Novoselov}}, \bibinfo {author}
  {\bibfnamefont {R.~V.}\ \bibnamefont {Gorbachev}}, \bibinfo {author}
  {\bibfnamefont {H.~S.}\ \bibnamefont {Shin}}, \bibinfo {author}
  {\bibfnamefont {V.~I.}\ \bibnamefont {Fal’ko}},\ and\ \bibinfo {author}
  {\bibfnamefont {A.~I.}\ \bibnamefont {Tartakovskii}},\ }\bibfield  {title}
  {\bibinfo {title} {Resonantly hybridized excitons in moir{\'e} superlattices
  in van der {W}aals heterostructures},\ }\href@noop {} {\bibfield  {journal}
  {\bibinfo  {journal} {Nature}\ }\textbf {\bibinfo {volume} {567}},\ \bibinfo
  {pages} {81} (\bibinfo {year} {2019})}\BibitemShut {NoStop}%
\bibitem [{\citenamefont {Hsu}\ \emph {et~al.}(2018)\citenamefont {Hsu},
  \citenamefont {Lu}, \citenamefont {Wu}, \citenamefont {Lee}, \citenamefont
  {Chen}, \citenamefont {Wu}, \citenamefont {Chou}, \citenamefont {Jeng},
  \citenamefont {Li}, \citenamefont {Chu},\ and\ \citenamefont
  {Chang}}]{hsu2018negative}%
  \BibitemOpen
  \bibfield  {author} {\bibinfo {author} {\bibfnamefont {W.-T.}\ \bibnamefont
  {Hsu}}, \bibinfo {author} {\bibfnamefont {L.-S.}\ \bibnamefont {Lu}},
  \bibinfo {author} {\bibfnamefont {P.-H.}\ \bibnamefont {Wu}}, \bibinfo
  {author} {\bibfnamefont {M.-H.}\ \bibnamefont {Lee}}, \bibinfo {author}
  {\bibfnamefont {P.-J.}\ \bibnamefont {Chen}}, \bibinfo {author}
  {\bibfnamefont {P.-Y.}\ \bibnamefont {Wu}}, \bibinfo {author} {\bibfnamefont
  {Y.-C.}\ \bibnamefont {Chou}}, \bibinfo {author} {\bibfnamefont {H.-T.}\
  \bibnamefont {Jeng}}, \bibinfo {author} {\bibfnamefont {L.-J.}\ \bibnamefont
  {Li}}, \bibinfo {author} {\bibfnamefont {M.-W.}\ \bibnamefont {Chu}},\ and\
  \bibinfo {author} {\bibfnamefont {W.-H.}\ \bibnamefont {Chang}},\ }\bibfield
  {title} {\bibinfo {title} {Negative circular polarization emissions from
  {WSe}$_2$/{MoSe}$_2$ commensurate heterobilayers},\ }\href@noop {} {\bibfield
   {journal} {\bibinfo  {journal} {Nat. Commun.}\ }\textbf {\bibinfo {volume}
  {9}},\ \bibinfo {pages} {1356} (\bibinfo {year} {2018})}\BibitemShut
  {NoStop}%
\bibitem [{\citenamefont {Choi}\ \emph {et~al.}(2020)\citenamefont {Choi},
  \citenamefont {Hsu}, \citenamefont {Lu}, \citenamefont {Sun}, \citenamefont
  {Cheng}, \citenamefont {Lee}, \citenamefont {Quan}, \citenamefont {Tran},
  \citenamefont {Wang}, \citenamefont {Staab}, \citenamefont {Jones},
  \citenamefont {Taniguchi}, \citenamefont {Watanabe}, \citenamefont {Chu},
  \citenamefont {Gwo}, \citenamefont {Kim}, \citenamefont {Shih}, \citenamefont
  {Li},\ and\ \citenamefont {Chang}}]{choi2020moire}%
  \BibitemOpen
  \bibfield  {author} {\bibinfo {author} {\bibfnamefont {J.}~\bibnamefont
  {Choi}}, \bibinfo {author} {\bibfnamefont {W.-T.}\ \bibnamefont {Hsu}},
  \bibinfo {author} {\bibfnamefont {L.-S.}\ \bibnamefont {Lu}}, \bibinfo
  {author} {\bibfnamefont {L.}~\bibnamefont {Sun}}, \bibinfo {author}
  {\bibfnamefont {H.-Y.}\ \bibnamefont {Cheng}}, \bibinfo {author}
  {\bibfnamefont {M.-H.}\ \bibnamefont {Lee}}, \bibinfo {author} {\bibfnamefont
  {J.}~\bibnamefont {Quan}}, \bibinfo {author} {\bibfnamefont {K.}~\bibnamefont
  {Tran}}, \bibinfo {author} {\bibfnamefont {C.-Y.}\ \bibnamefont {Wang}},
  \bibinfo {author} {\bibfnamefont {M.}~\bibnamefont {Staab}}, \bibinfo
  {author} {\bibfnamefont {K.}~\bibnamefont {Jones}}, \bibinfo {author}
  {\bibfnamefont {T.}~\bibnamefont {Taniguchi}}, \bibinfo {author}
  {\bibfnamefont {K.}~\bibnamefont {Watanabe}}, \bibinfo {author}
  {\bibfnamefont {M.-W.}\ \bibnamefont {Chu}}, \bibinfo {author} {\bibfnamefont
  {S.}~\bibnamefont {Gwo}}, \bibinfo {author} {\bibfnamefont {S.}~\bibnamefont
  {Kim}}, \bibinfo {author} {\bibfnamefont {C.-K.}\ \bibnamefont {Shih}},
  \bibinfo {author} {\bibfnamefont {X.}~\bibnamefont {Li}},\ and\ \bibinfo
  {author} {\bibfnamefont {W.-H.}\ \bibnamefont {Chang}},\ }\bibfield  {title}
  {\bibinfo {title} {Moir{\'e} potential impedes interlayer exciton diffusion
  in van der {W}aals heterostructures},\ }\href@noop {} {\bibfield  {journal}
  {\bibinfo  {journal} {Sci. Adv.}\ }\textbf {\bibinfo {volume} {6}},\ \bibinfo
  {pages} {eaba8866} (\bibinfo {year} {2020})}\BibitemShut {NoStop}%
\bibitem [{\citenamefont {Xia}\ \emph {et~al.}(2021)\citenamefont {Xia},
  \citenamefont {Yan}, \citenamefont {Wang}, \citenamefont {He}, \citenamefont
  {Gong}, \citenamefont {Chen}, \citenamefont {Sum}, \citenamefont {Liu},
  \citenamefont {Ajayan},\ and\ \citenamefont {Shen}}]{xia2021strong}%
  \BibitemOpen
  \bibfield  {author} {\bibinfo {author} {\bibfnamefont {J.}~\bibnamefont
  {Xia}}, \bibinfo {author} {\bibfnamefont {J.}~\bibnamefont {Yan}}, \bibinfo
  {author} {\bibfnamefont {Z.}~\bibnamefont {Wang}}, \bibinfo {author}
  {\bibfnamefont {Y.}~\bibnamefont {He}}, \bibinfo {author} {\bibfnamefont
  {Y.}~\bibnamefont {Gong}}, \bibinfo {author} {\bibfnamefont {W.}~\bibnamefont
  {Chen}}, \bibinfo {author} {\bibfnamefont {T.~C.}\ \bibnamefont {Sum}},
  \bibinfo {author} {\bibfnamefont {Z.}~\bibnamefont {Liu}}, \bibinfo {author}
  {\bibfnamefont {P.~M.}\ \bibnamefont {Ajayan}},\ and\ \bibinfo {author}
  {\bibfnamefont {Z.}~\bibnamefont {Shen}},\ }\bibfield  {title} {\bibinfo
  {title} {Strong coupling and pressure engineering in {WSe$_2$-MoSe$_2$}
  heterobilayers},\ }\href@noop {} {\bibfield  {journal} {\bibinfo  {journal}
  {Nat. Phys.}\ }\textbf {\bibinfo {volume} {17}},\ \bibinfo {pages} {92}
  (\bibinfo {year} {2021})}\BibitemShut {NoStop}%
\bibitem [{\citenamefont {Butov}(2017)}]{Butov2017}%
  \BibitemOpen
  \bibfield  {author} {\bibinfo {author} {\bibfnamefont {L.~V.}\ \bibnamefont
  {Butov}},\ }\bibfield  {title} {\bibinfo {title} {Excitonic devices},\
  }\href@noop {} {\bibfield  {journal} {\bibinfo  {journal} {Superlattices and
  Microstructures}\ }\textbf {\bibinfo {volume} {108}},\ \bibinfo {pages} {2}
  (\bibinfo {year} {2017})}\BibitemShut {NoStop}%
\bibitem [{\citenamefont {Sun}\ \emph {et~al.}(2022)\citenamefont {Sun},
  \citenamefont {Ciarrocchi}, \citenamefont {Tagarelli}, \citenamefont
  {Gonzalez~Marin}, \citenamefont {Watanabe}, \citenamefont {Taniguchi},\ and\
  \citenamefont {Kis}}]{Sun2022}%
  \BibitemOpen
  \bibfield  {author} {\bibinfo {author} {\bibfnamefont {Z.}~\bibnamefont
  {Sun}}, \bibinfo {author} {\bibfnamefont {A.}~\bibnamefont {Ciarrocchi}},
  \bibinfo {author} {\bibfnamefont {F.}~\bibnamefont {Tagarelli}}, \bibinfo
  {author} {\bibfnamefont {J.~F.}\ \bibnamefont {Gonzalez~Marin}}, \bibinfo
  {author} {\bibfnamefont {K.}~\bibnamefont {Watanabe}}, \bibinfo {author}
  {\bibfnamefont {T.}~\bibnamefont {Taniguchi}},\ and\ \bibinfo {author}
  {\bibfnamefont {A.}~\bibnamefont {Kis}},\ }\bibfield  {title} {\bibinfo
  {title} {{Excitonic transport driven by repulsive dipolar interaction in a
  van der Waals heterostructure}},\ }\href@noop {} {\bibfield  {journal}
  {\bibinfo  {journal} {Nat. Photon.}\ }\textbf {\bibinfo {volume} {16}},\
  \bibinfo {pages} {79} (\bibinfo {year} {2022})}\BibitemShut {NoStop}%
\bibitem [{\citenamefont {Shanks}\ \emph
  {et~al.}(2022{\natexlab{a}})\citenamefont {Shanks}, \citenamefont
  {Mahdikhanysarvejahany}, \citenamefont {Stanfill}, \citenamefont {Koehler},
  \citenamefont {Mandrus}, \citenamefont {Taniguchi}, \citenamefont {Watanabe},
  \citenamefont {LeRoy},\ and\ \citenamefont {Schaibley}}]{Shanks2022}%
  \BibitemOpen
  \bibfield  {author} {\bibinfo {author} {\bibfnamefont {D.~N.}\ \bibnamefont
  {Shanks}}, \bibinfo {author} {\bibfnamefont {F.}~\bibnamefont
  {Mahdikhanysarvejahany}}, \bibinfo {author} {\bibfnamefont {T.~G.}\
  \bibnamefont {Stanfill}}, \bibinfo {author} {\bibfnamefont {M.~R.}\
  \bibnamefont {Koehler}}, \bibinfo {author} {\bibfnamefont {D.~G.}\
  \bibnamefont {Mandrus}}, \bibinfo {author} {\bibfnamefont {T.}~\bibnamefont
  {Taniguchi}}, \bibinfo {author} {\bibfnamefont {K.}~\bibnamefont {Watanabe}},
  \bibinfo {author} {\bibfnamefont {B.~J.}\ \bibnamefont {LeRoy}},\ and\
  \bibinfo {author} {\bibfnamefont {J.~R.}\ \bibnamefont {Schaibley}},\
  }\bibfield  {title} {\bibinfo {title} {Interlayer exciton diode and
  transistor},\ }\href@noop {} {\bibfield  {journal} {\bibinfo  {journal} {Nano
  Lett.}\ }\textbf {\bibinfo {volume} {22}},\ \bibinfo {pages} {6599} (\bibinfo
  {year} {2022}{\natexlab{a}})}\BibitemShut {NoStop}%
\bibitem [{\citenamefont {Yuan}\ \emph {et~al.}(2020)\citenamefont {Yuan},
  \citenamefont {Zheng}, \citenamefont {Kunstmann}, \citenamefont {Brumme},
  \citenamefont {Kuc}, \citenamefont {Ma}, \citenamefont {Deng}, \citenamefont
  {Blach}, \citenamefont {Pan},\ and\ \citenamefont {Huang}}]{Yuan2020}%
  \BibitemOpen
  \bibfield  {author} {\bibinfo {author} {\bibfnamefont {L.}~\bibnamefont
  {Yuan}}, \bibinfo {author} {\bibfnamefont {B.}~\bibnamefont {Zheng}},
  \bibinfo {author} {\bibfnamefont {J.}~\bibnamefont {Kunstmann}}, \bibinfo
  {author} {\bibfnamefont {T.}~\bibnamefont {Brumme}}, \bibinfo {author}
  {\bibfnamefont {A.~B.}\ \bibnamefont {Kuc}}, \bibinfo {author} {\bibfnamefont
  {C.}~\bibnamefont {Ma}}, \bibinfo {author} {\bibfnamefont {S.}~\bibnamefont
  {Deng}}, \bibinfo {author} {\bibfnamefont {D.}~\bibnamefont {Blach}},
  \bibinfo {author} {\bibfnamefont {A.}~\bibnamefont {Pan}},\ and\ \bibinfo
  {author} {\bibfnamefont {L.}~\bibnamefont {Huang}},\ }\bibfield  {title}
  {\bibinfo {title} {{Twist-angle-dependent interlayer exciton diffusion in
  WS$_2$-WSe$_2$ heterobilayers}},\ }\href@noop {} {\bibfield  {journal}
  {\bibinfo  {journal} {Nat. Mater.}\ }\textbf {\bibinfo {volume} {19}},\
  \bibinfo {pages} {617} (\bibinfo {year} {2020})}\BibitemShut {NoStop}%
\bibitem [{\citenamefont {Wang}\ \emph {et~al.}(2021)\citenamefont {Wang},
  \citenamefont {Shi}, \citenamefont {Shih}, \citenamefont {Zhou},
  \citenamefont {Wu}, \citenamefont {Bai}, \citenamefont {Rhodes},
  \citenamefont {Barmak}, \citenamefont {Hone}, \citenamefont {Dean},\ and\
  \citenamefont {Zhu}}]{Wang2021}%
  \BibitemOpen
  \bibfield  {author} {\bibinfo {author} {\bibfnamefont {J.}~\bibnamefont
  {Wang}}, \bibinfo {author} {\bibfnamefont {Q.}~\bibnamefont {Shi}}, \bibinfo
  {author} {\bibfnamefont {E.-M.}\ \bibnamefont {Shih}}, \bibinfo {author}
  {\bibfnamefont {L.}~\bibnamefont {Zhou}}, \bibinfo {author} {\bibfnamefont
  {W.}~\bibnamefont {Wu}}, \bibinfo {author} {\bibfnamefont {Y.}~\bibnamefont
  {Bai}}, \bibinfo {author} {\bibfnamefont {D.}~\bibnamefont {Rhodes}},
  \bibinfo {author} {\bibfnamefont {K.}~\bibnamefont {Barmak}}, \bibinfo
  {author} {\bibfnamefont {J.}~\bibnamefont {Hone}}, \bibinfo {author}
  {\bibfnamefont {C.~R.}\ \bibnamefont {Dean}},\ and\ \bibinfo {author}
  {\bibfnamefont {X.-Y.}\ \bibnamefont {Zhu}},\ }\bibfield  {title} {\bibinfo
  {title} {{Diffusivity Reveals Three Distinct Phases of Interlayer Excitons in
  ${\mathrm{MoSe}}_{2}/{\mathrm{WSe}}_{2}$ Heterobilayers}},\ }\href@noop {}
  {\bibfield  {journal} {\bibinfo  {journal} {Phys. Rev. Lett.}\ }\textbf
  {\bibinfo {volume} {126}},\ \bibinfo {pages} {106804} (\bibinfo {year}
  {2021})}\BibitemShut {NoStop}%
\bibitem [{\citenamefont {Jiang}\ \emph {et~al.}(2021)\citenamefont {Jiang},
  \citenamefont {Chen}, \citenamefont {Zheng}, \citenamefont {Zheng},\ and\
  \citenamefont {Pan}}]{Jiang2021}%
  \BibitemOpen
  \bibfield  {author} {\bibinfo {author} {\bibfnamefont {Y.}~\bibnamefont
  {Jiang}}, \bibinfo {author} {\bibfnamefont {S.}~\bibnamefont {Chen}},
  \bibinfo {author} {\bibfnamefont {W.}~\bibnamefont {Zheng}}, \bibinfo
  {author} {\bibfnamefont {B.}~\bibnamefont {Zheng}},\ and\ \bibinfo {author}
  {\bibfnamefont {A.}~\bibnamefont {Pan}},\ }\bibfield  {title} {\bibinfo
  {title} {{Interlayer exciton formation, relaxation, and transport in TMD van
  der Waals heterostructures}},\ }\href@noop {} {\bibfield  {journal} {\bibinfo
   {journal} {Light}\ }\textbf {\bibinfo {volume} {10}},\ \bibinfo {pages} {72}
  (\bibinfo {year} {2021})}\BibitemShut {NoStop}%
\bibitem [{\citenamefont {Ciarrocchi}\ \emph {et~al.}(2022)\citenamefont
  {Ciarrocchi}, \citenamefont {Tagarelli}, \citenamefont {Avsar},\ and\
  \citenamefont {Kis}}]{Ciarrocchi2022}%
  \BibitemOpen
  \bibfield  {author} {\bibinfo {author} {\bibfnamefont {A.}~\bibnamefont
  {Ciarrocchi}}, \bibinfo {author} {\bibfnamefont {F.}~\bibnamefont
  {Tagarelli}}, \bibinfo {author} {\bibfnamefont {A.}~\bibnamefont {Avsar}},\
  and\ \bibinfo {author} {\bibfnamefont {A.}~\bibnamefont {Kis}},\ }\bibfield
  {title} {\bibinfo {title} {{Excitonic devices with van der Waals
  heterostructures: valleytronics meets twistronics}},\ }\href@noop {}
  {\bibfield  {journal} {\bibinfo  {journal} {Nat. Rev. Mater.}\ }\textbf
  {\bibinfo {volume} {7}},\ \bibinfo {pages} {449} (\bibinfo {year}
  {2022})}\BibitemShut {NoStop}%
\bibitem [{\citenamefont {Shanks}\ \emph {et~al.}(2021)\citenamefont {Shanks},
  \citenamefont {Mahdikhanysarvejahany}, \citenamefont {Muccianti},
  \citenamefont {Alfrey}, \citenamefont {Koehler}, \citenamefont {Mandrus},
  \citenamefont {Taniguchi}, \citenamefont {Watanabe}, \citenamefont {Yu},
  \citenamefont {LeRoy},\ and\ \citenamefont
  {Schaibley}}]{shanks2021nanoscale}%
  \BibitemOpen
  \bibfield  {author} {\bibinfo {author} {\bibfnamefont {D.~N.}\ \bibnamefont
  {Shanks}}, \bibinfo {author} {\bibfnamefont {F.}~\bibnamefont
  {Mahdikhanysarvejahany}}, \bibinfo {author} {\bibfnamefont {C.}~\bibnamefont
  {Muccianti}}, \bibinfo {author} {\bibfnamefont {A.}~\bibnamefont {Alfrey}},
  \bibinfo {author} {\bibfnamefont {M.~R.}\ \bibnamefont {Koehler}}, \bibinfo
  {author} {\bibfnamefont {D.~G.}\ \bibnamefont {Mandrus}}, \bibinfo {author}
  {\bibfnamefont {T.}~\bibnamefont {Taniguchi}}, \bibinfo {author}
  {\bibfnamefont {K.}~\bibnamefont {Watanabe}}, \bibinfo {author}
  {\bibfnamefont {H.}~\bibnamefont {Yu}}, \bibinfo {author} {\bibfnamefont
  {B.~J.}\ \bibnamefont {LeRoy}},\ and\ \bibinfo {author} {\bibfnamefont
  {J.~R.}\ \bibnamefont {Schaibley}},\ }\bibfield  {title} {\bibinfo {title}
  {Nanoscale trapping of interlayer excitons in a {2D} semiconductor
  heterostructure},\ }\href@noop {} {\bibfield  {journal} {\bibinfo  {journal}
  {Nano Lett.}\ }\textbf {\bibinfo {volume} {21}},\ \bibinfo {pages} {5641}
  (\bibinfo {year} {2021})}\BibitemShut {NoStop}%
\bibitem [{\citenamefont {Shanks}\ \emph
  {et~al.}(2022{\natexlab{b}})\citenamefont {Shanks}, \citenamefont
  {Mahdikhanysarvejahany}, \citenamefont {Koehler}, \citenamefont {Mandrus},
  \citenamefont {Taniguchi}, \citenamefont {Watanabe}, \citenamefont {LeRoy},\
  and\ \citenamefont {Schaibley}}]{shanks2022single}%
  \BibitemOpen
  \bibfield  {author} {\bibinfo {author} {\bibfnamefont {D.~N.}\ \bibnamefont
  {Shanks}}, \bibinfo {author} {\bibfnamefont {F.}~\bibnamefont
  {Mahdikhanysarvejahany}}, \bibinfo {author} {\bibfnamefont {M.~R.}\
  \bibnamefont {Koehler}}, \bibinfo {author} {\bibfnamefont {D.~G.}\
  \bibnamefont {Mandrus}}, \bibinfo {author} {\bibfnamefont {T.}~\bibnamefont
  {Taniguchi}}, \bibinfo {author} {\bibfnamefont {K.}~\bibnamefont {Watanabe}},
  \bibinfo {author} {\bibfnamefont {B.~J.}\ \bibnamefont {LeRoy}},\ and\
  \bibinfo {author} {\bibfnamefont {J.~R.}\ \bibnamefont {Schaibley}},\
  }\bibfield  {title} {\bibinfo {title} {Single exciton trapping in an
  electrostatically defined {2D} semiconductor quantum dot},\ }\href@noop {}
  {\bibfield  {journal} {\bibinfo  {journal} {Phys. Rev. B}\ }\textbf {\bibinfo
  {volume} {106}},\ \bibinfo {pages} {L201401} (\bibinfo {year}
  {2022}{\natexlab{b}})}\BibitemShut {NoStop}%
\bibitem [{\citenamefont {Shi}\ \emph {et~al.}(2019)\citenamefont {Shi},
  \citenamefont {Ma},\ and\ \citenamefont {Song}}]{shi2019gate}%
  \BibitemOpen
  \bibfield  {author} {\bibinfo {author} {\bibfnamefont {L.-k.}\ \bibnamefont
  {Shi}}, \bibinfo {author} {\bibfnamefont {J.}~\bibnamefont {Ma}},\ and\
  \bibinfo {author} {\bibfnamefont {J.~C.}\ \bibnamefont {Song}},\ }\bibfield
  {title} {\bibinfo {title} {Gate-tunable flat bands in van der {W}aals
  patterned dielectric superlattices},\ }\href@noop {} {\bibfield  {journal}
  {\bibinfo  {journal} {2D Mater.}\ }\textbf {\bibinfo {volume} {7}},\ \bibinfo
  {pages} {015028} (\bibinfo {year} {2019})}\BibitemShut {NoStop}%
\bibitem [{\citenamefont {Carr}\ \emph {et~al.}(2018)\citenamefont {Carr},
  \citenamefont {Massatt}, \citenamefont {Torrisi}, \citenamefont {Cazeaux},
  \citenamefont {Luskin},\ and\ \citenamefont {Kaxiras}}]{carr2018relaxation}%
  \BibitemOpen
  \bibfield  {author} {\bibinfo {author} {\bibfnamefont {S.}~\bibnamefont
  {Carr}}, \bibinfo {author} {\bibfnamefont {D.}~\bibnamefont {Massatt}},
  \bibinfo {author} {\bibfnamefont {S.~B.}\ \bibnamefont {Torrisi}}, \bibinfo
  {author} {\bibfnamefont {P.}~\bibnamefont {Cazeaux}}, \bibinfo {author}
  {\bibfnamefont {M.}~\bibnamefont {Luskin}},\ and\ \bibinfo {author}
  {\bibfnamefont {E.}~\bibnamefont {Kaxiras}},\ }\bibfield  {title} {\bibinfo
  {title} {Relaxation and domain formation in incommensurate two-dimensional
  heterostructures},\ }\href@noop {} {\bibfield  {journal} {\bibinfo  {journal}
  {Phys. Rev. B}\ }\textbf {\bibinfo {volume} {98}},\ \bibinfo {pages} {224102}
  (\bibinfo {year} {2018})}\BibitemShut {NoStop}%
\bibitem [{\citenamefont {Enaldiev}\ \emph {et~al.}(2020)\citenamefont
  {Enaldiev}, \citenamefont {Z{\'o}lyomi}, \citenamefont {Yelgel},
  \citenamefont {Magorrian},\ and\ \citenamefont
  {Fal’ko}}]{enaldiev2020stacking}%
  \BibitemOpen
  \bibfield  {author} {\bibinfo {author} {\bibfnamefont {V.}~\bibnamefont
  {Enaldiev}}, \bibinfo {author} {\bibfnamefont {V.}~\bibnamefont
  {Z{\'o}lyomi}}, \bibinfo {author} {\bibfnamefont {C.}~\bibnamefont {Yelgel}},
  \bibinfo {author} {\bibfnamefont {S.}~\bibnamefont {Magorrian}},\ and\
  \bibinfo {author} {\bibfnamefont {V.}~\bibnamefont {Fal’ko}},\ }\bibfield
  {title} {\bibinfo {title} {Stacking domains and dislocation networks in
  marginally twisted bilayers of transition metal dichalcogenides},\
  }\href@noop {} {\bibfield  {journal} {\bibinfo  {journal} {Phys. Rev. Lett.}\
  }\textbf {\bibinfo {volume} {124}},\ \bibinfo {pages} {206101} (\bibinfo
  {year} {2020})}\BibitemShut {NoStop}%
\bibitem [{\citenamefont {Rosenberger}\ \emph {et~al.}(2020)\citenamefont
  {Rosenberger}, \citenamefont {Chuang}, \citenamefont {Phillips},
  \citenamefont {Oleshko}, \citenamefont {McCreary}, \citenamefont {Sivaram},
  \citenamefont {Hellberg},\ and\ \citenamefont
  {Jonker}}]{rosenberger2020twist}%
  \BibitemOpen
  \bibfield  {author} {\bibinfo {author} {\bibfnamefont {M.~R.}\ \bibnamefont
  {Rosenberger}}, \bibinfo {author} {\bibfnamefont {H.-J.}\ \bibnamefont
  {Chuang}}, \bibinfo {author} {\bibfnamefont {M.}~\bibnamefont {Phillips}},
  \bibinfo {author} {\bibfnamefont {V.~P.}\ \bibnamefont {Oleshko}}, \bibinfo
  {author} {\bibfnamefont {K.~M.}\ \bibnamefont {McCreary}}, \bibinfo {author}
  {\bibfnamefont {S.~V.}\ \bibnamefont {Sivaram}}, \bibinfo {author}
  {\bibfnamefont {C.~S.}\ \bibnamefont {Hellberg}},\ and\ \bibinfo {author}
  {\bibfnamefont {B.~T.}\ \bibnamefont {Jonker}},\ }\bibfield  {title}
  {\bibinfo {title} {Twist angle-dependent atomic reconstruction and moir{\'e}
  patterns in transition metal dichalcogenide heterostructures},\ }\href@noop
  {} {\bibfield  {journal} {\bibinfo  {journal} {ACS Nano}\ }\textbf {\bibinfo
  {volume} {14}},\ \bibinfo {pages} {4550} (\bibinfo {year}
  {2020})}\BibitemShut {NoStop}%
\bibitem [{\citenamefont {Zhao}\ \emph {et~al.}(2022)\citenamefont {Zhao},
  \citenamefont {Li}, \citenamefont {Huang}, \citenamefont {Rupp},
  \citenamefont {G{\"o}ser}, \citenamefont {Vovk}, \citenamefont {Kruchinin},
  \citenamefont {Watanabe}, \citenamefont {Taniguchi}, \citenamefont {Bilgin},
  \citenamefont {Baimuratov},\ and\ \citenamefont
  {Högele}}]{zhao2022excitons}%
  \BibitemOpen
  \bibfield  {author} {\bibinfo {author} {\bibfnamefont {S.}~\bibnamefont
  {Zhao}}, \bibinfo {author} {\bibfnamefont {Z.}~\bibnamefont {Li}}, \bibinfo
  {author} {\bibfnamefont {X.}~\bibnamefont {Huang}}, \bibinfo {author}
  {\bibfnamefont {A.}~\bibnamefont {Rupp}}, \bibinfo {author} {\bibfnamefont
  {J.}~\bibnamefont {G{\"o}ser}}, \bibinfo {author} {\bibfnamefont {I.~A.}\
  \bibnamefont {Vovk}}, \bibinfo {author} {\bibfnamefont {S.~Y.}\ \bibnamefont
  {Kruchinin}}, \bibinfo {author} {\bibfnamefont {K.}~\bibnamefont {Watanabe}},
  \bibinfo {author} {\bibfnamefont {T.}~\bibnamefont {Taniguchi}}, \bibinfo
  {author} {\bibfnamefont {I.}~\bibnamefont {Bilgin}}, \bibinfo {author}
  {\bibfnamefont {A.~S.}\ \bibnamefont {Baimuratov}},\ and\ \bibinfo {author}
  {\bibfnamefont {A.}~\bibnamefont {Högele}},\ }\bibfield  {title} {\bibinfo
  {title} {Excitons in mesoscopically reconstructed moir{\'e}
  heterostructures},\ }\href@noop {} {\bibfield  {journal} {\bibinfo  {journal}
  {2022, arXiv:2202.11139}\ ,\ \bibinfo {pages} {arXiv preprint
  https://arxiv.org/abs/2202.11139}} (\bibinfo {year} {accessed March 3,
  2022})}\BibitemShut {NoStop}%
\bibitem [{\citenamefont {Ashida}\ \emph {et~al.}(2015)\citenamefont {Ashida},
  \citenamefont {Kajino}, \citenamefont {Kutsuma}, \citenamefont {Ohtani},\
  and\ \citenamefont {Kaneko}}]{ashida2015crystallographic}%
  \BibitemOpen
  \bibfield  {author} {\bibinfo {author} {\bibfnamefont {K.}~\bibnamefont
  {Ashida}}, \bibinfo {author} {\bibfnamefont {T.}~\bibnamefont {Kajino}},
  \bibinfo {author} {\bibfnamefont {Y.}~\bibnamefont {Kutsuma}}, \bibinfo
  {author} {\bibfnamefont {N.}~\bibnamefont {Ohtani}},\ and\ \bibinfo {author}
  {\bibfnamefont {T.}~\bibnamefont {Kaneko}},\ }\bibfield  {title} {\bibinfo
  {title} {Crystallographic orientation dependence of {SEM} contrast revealed
  by {SiC} polytypes},\ }\href@noop {} {\bibfield  {journal} {\bibinfo
  {journal} {J. Vac. Sci. Technol. B}\ }\textbf {\bibinfo {volume} {33}},\
  \bibinfo {pages} {04E104} (\bibinfo {year} {2015})}\BibitemShut {NoStop}%
\bibitem [{\citenamefont {Andersen}\ \emph {et~al.}(2021)\citenamefont
  {Andersen}, \citenamefont {Scuri}, \citenamefont {Sushko}, \citenamefont
  {De~Greve}, \citenamefont {Sung}, \citenamefont {Zhou}, \citenamefont {Wild},
  \citenamefont {Gelly}, \citenamefont {Heo}, \citenamefont {B{\'e}rub{\'e}},
  \citenamefont {Joe}, \citenamefont {Jauregui}, \citenamefont {Watanabe},
  \citenamefont {Taniguchi}, \citenamefont {Kim}, \citenamefont {Park},\ and\
  \citenamefont {Lukin}}]{andersen2021excitons}%
  \BibitemOpen
  \bibfield  {author} {\bibinfo {author} {\bibfnamefont {T.~I.}\ \bibnamefont
  {Andersen}}, \bibinfo {author} {\bibfnamefont {G.}~\bibnamefont {Scuri}},
  \bibinfo {author} {\bibfnamefont {A.}~\bibnamefont {Sushko}}, \bibinfo
  {author} {\bibfnamefont {K.}~\bibnamefont {De~Greve}}, \bibinfo {author}
  {\bibfnamefont {J.}~\bibnamefont {Sung}}, \bibinfo {author} {\bibfnamefont
  {Y.}~\bibnamefont {Zhou}}, \bibinfo {author} {\bibfnamefont {D.~S.}\
  \bibnamefont {Wild}}, \bibinfo {author} {\bibfnamefont {R.~J.}\ \bibnamefont
  {Gelly}}, \bibinfo {author} {\bibfnamefont {H.}~\bibnamefont {Heo}}, \bibinfo
  {author} {\bibfnamefont {D.}~\bibnamefont {B{\'e}rub{\'e}}}, \bibinfo
  {author} {\bibfnamefont {A.~Y.}\ \bibnamefont {Joe}}, \bibinfo {author}
  {\bibfnamefont {L.~A.}\ \bibnamefont {Jauregui}}, \bibinfo {author}
  {\bibfnamefont {K.}~\bibnamefont {Watanabe}}, \bibinfo {author}
  {\bibfnamefont {T.}~\bibnamefont {Taniguchi}}, \bibinfo {author}
  {\bibfnamefont {P.}~\bibnamefont {Kim}}, \bibinfo {author} {\bibfnamefont
  {H.}~\bibnamefont {Park}},\ and\ \bibinfo {author} {\bibfnamefont {M.~D.}\
  \bibnamefont {Lukin}},\ }\bibfield  {title} {\bibinfo {title} {Excitons in a
  reconstructed moir{\'e} potential in twisted {WSe}$_2$/{WSe}$_2$
  homobilayers},\ }\href@noop {} {\bibfield  {journal} {\bibinfo  {journal}
  {Nat. Mater.}\ }\textbf {\bibinfo {volume} {20}},\ \bibinfo {pages} {480}
  (\bibinfo {year} {2021})}\BibitemShut {NoStop}%
\bibitem [{\citenamefont {Yu}\ \emph {et~al.}(2017)\citenamefont {Yu},
  \citenamefont {Liu}, \citenamefont {Tang}, \citenamefont {Xu},\ and\
  \citenamefont {Yao}}]{yu2017moire}%
  \BibitemOpen
  \bibfield  {author} {\bibinfo {author} {\bibfnamefont {H.}~\bibnamefont
  {Yu}}, \bibinfo {author} {\bibfnamefont {G.-B.}\ \bibnamefont {Liu}},
  \bibinfo {author} {\bibfnamefont {J.}~\bibnamefont {Tang}}, \bibinfo {author}
  {\bibfnamefont {X.}~\bibnamefont {Xu}},\ and\ \bibinfo {author}
  {\bibfnamefont {W.}~\bibnamefont {Yao}},\ }\bibfield  {title} {\bibinfo
  {title} {Moir{\'e} excitons: From programmable quantum emitter arrays to
  spin-orbit--coupled artificial lattices},\ }\href@noop {} {\bibfield
  {journal} {\bibinfo  {journal} {Sci. Adv.}\ }\textbf {\bibinfo {volume}
  {3}},\ \bibinfo {pages} {e1701696} (\bibinfo {year} {2017})}\BibitemShut
  {NoStop}%
\bibitem [{\citenamefont {Wo{\'z}niak}\ \emph {et~al.}(2020)\citenamefont
  {Wo{\'z}niak}, \citenamefont {Junior}, \citenamefont {Seifert}, \citenamefont
  {Chaves},\ and\ \citenamefont {Kunstmann}}]{wozniak2020exciton}%
  \BibitemOpen
  \bibfield  {author} {\bibinfo {author} {\bibfnamefont {T.}~\bibnamefont
  {Wo{\'z}niak}}, \bibinfo {author} {\bibfnamefont {P.~E.~F.}\ \bibnamefont
  {Junior}}, \bibinfo {author} {\bibfnamefont {G.}~\bibnamefont {Seifert}},
  \bibinfo {author} {\bibfnamefont {A.}~\bibnamefont {Chaves}},\ and\ \bibinfo
  {author} {\bibfnamefont {J.}~\bibnamefont {Kunstmann}},\ }\bibfield  {title}
  {\bibinfo {title} {Exciton g-factors of van der {W}aals heterostructures from
  first principles calculations},\ }\href@noop {} {\bibfield  {journal}
  {\bibinfo  {journal} {Phys. Rev. B}\ }\textbf {\bibinfo {volume} {101}},\
  \bibinfo {pages} {235408} (\bibinfo {year} {2020})}\BibitemShut {NoStop}%
\bibitem [{\citenamefont {Wu}\ \emph {et~al.}(2018{\natexlab{b}})\citenamefont
  {Wu}, \citenamefont {Lovorn},\ and\ \citenamefont
  {MacDonald}}]{wu2018theory}%
  \BibitemOpen
  \bibfield  {author} {\bibinfo {author} {\bibfnamefont {F.}~\bibnamefont
  {Wu}}, \bibinfo {author} {\bibfnamefont {T.}~\bibnamefont {Lovorn}},\ and\
  \bibinfo {author} {\bibfnamefont {A.}~\bibnamefont {MacDonald}},\ }\bibfield
  {title} {\bibinfo {title} {Theory of optical absorption by interlayer
  excitons in transition metal dichalcogenide heterobilayers},\ }\href@noop {}
  {\bibfield  {journal} {\bibinfo  {journal} {Phys. Rev. B}\ }\textbf {\bibinfo
  {volume} {97}},\ \bibinfo {pages} {035306} (\bibinfo {year}
  {2018}{\natexlab{b}})}\BibitemShut {NoStop}%
\bibitem [{\citenamefont {F{\"o}rg}\ \emph {et~al.}(2019)\citenamefont
  {F{\"o}rg}, \citenamefont {Colombier}, \citenamefont {Patel}, \citenamefont
  {Lindlau}, \citenamefont {Mohite}, \citenamefont {Yamaguchi}, \citenamefont
  {Glazov}, \citenamefont {Hunger},\ and\ \citenamefont
  {H{\"o}gele}}]{forg2019cavity}%
  \BibitemOpen
  \bibfield  {author} {\bibinfo {author} {\bibfnamefont {M.}~\bibnamefont
  {F{\"o}rg}}, \bibinfo {author} {\bibfnamefont {L.}~\bibnamefont {Colombier}},
  \bibinfo {author} {\bibfnamefont {R.~K.}\ \bibnamefont {Patel}}, \bibinfo
  {author} {\bibfnamefont {J.}~\bibnamefont {Lindlau}}, \bibinfo {author}
  {\bibfnamefont {A.~D.}\ \bibnamefont {Mohite}}, \bibinfo {author}
  {\bibfnamefont {H.}~\bibnamefont {Yamaguchi}}, \bibinfo {author}
  {\bibfnamefont {M.~M.}\ \bibnamefont {Glazov}}, \bibinfo {author}
  {\bibfnamefont {D.}~\bibnamefont {Hunger}},\ and\ \bibinfo {author}
  {\bibfnamefont {A.}~\bibnamefont {H{\"o}gele}},\ }\bibfield  {title}
  {\bibinfo {title} {Cavity-control of interlayer excitons in van der {W}aals
  heterostructures},\ }\href@noop {} {\bibfield  {journal} {\bibinfo  {journal}
  {Nat. Commun.}\ }\textbf {\bibinfo {volume} {10}},\ \bibinfo {pages} {3697}
  (\bibinfo {year} {2019})}\BibitemShut {NoStop}%
\bibitem [{\citenamefont {Gillen}\ and\ \citenamefont
  {Maultzsch}(2018)}]{gillen2018interlayer}%
  \BibitemOpen
  \bibfield  {author} {\bibinfo {author} {\bibfnamefont {R.}~\bibnamefont
  {Gillen}}\ and\ \bibinfo {author} {\bibfnamefont {J.}~\bibnamefont
  {Maultzsch}},\ }\bibfield  {title} {\bibinfo {title} {Interlayer excitons in
  {MoSe}$_2$/{WSe}$_2$ heterostructures from first principles},\ }\href@noop {}
  {\bibfield  {journal} {\bibinfo  {journal} {Phys. Rev. B}\ }\textbf {\bibinfo
  {volume} {97}},\ \bibinfo {pages} {165306} (\bibinfo {year}
  {2018})}\BibitemShut {NoStop}%
\bibitem [{\citenamefont {F{\"o}rg}\ \emph {et~al.}(2021)\citenamefont
  {F{\"o}rg}, \citenamefont {Baimuratov}, \citenamefont {Kruchinin},
  \citenamefont {Vovk}, \citenamefont {Scherzer}, \citenamefont {F{\"o}rste},
  \citenamefont {Funk}, \citenamefont {Watanabe}, \citenamefont {Taniguchi},\
  and\ \citenamefont {H{\"o}gele}}]{forg2021moire}%
  \BibitemOpen
  \bibfield  {author} {\bibinfo {author} {\bibfnamefont {M.}~\bibnamefont
  {F{\"o}rg}}, \bibinfo {author} {\bibfnamefont {A.~S.}\ \bibnamefont
  {Baimuratov}}, \bibinfo {author} {\bibfnamefont {S.~Y.}\ \bibnamefont
  {Kruchinin}}, \bibinfo {author} {\bibfnamefont {I.~A.}\ \bibnamefont {Vovk}},
  \bibinfo {author} {\bibfnamefont {J.}~\bibnamefont {Scherzer}}, \bibinfo
  {author} {\bibfnamefont {J.}~\bibnamefont {F{\"o}rste}}, \bibinfo {author}
  {\bibfnamefont {V.}~\bibnamefont {Funk}}, \bibinfo {author} {\bibfnamefont
  {K.}~\bibnamefont {Watanabe}}, \bibinfo {author} {\bibfnamefont
  {T.}~\bibnamefont {Taniguchi}},\ and\ \bibinfo {author} {\bibfnamefont
  {A.}~\bibnamefont {H{\"o}gele}},\ }\bibfield  {title} {\bibinfo {title}
  {Moir{\'e} excitons in {MoSe}$_2$-{WSe}$_2$ heterobilayers and
  heterotrilayers},\ }\href@noop {} {\bibfield  {journal} {\bibinfo  {journal}
  {Nat. Commun.}\ }\textbf {\bibinfo {volume} {12}},\ \bibinfo {pages} {1656}
  (\bibinfo {year} {2021})}\BibitemShut {NoStop}%
\bibitem [{\citenamefont {Hanbicki}\ \emph {et~al.}(2018)\citenamefont
  {Hanbicki}, \citenamefont {Chuang}, \citenamefont {Rosenberger},
  \citenamefont {Hellberg}, \citenamefont {Sivaram}, \citenamefont {McCreary},
  \citenamefont {Mazin},\ and\ \citenamefont {Jonker}}]{hanbicki2018double}%
  \BibitemOpen
  \bibfield  {author} {\bibinfo {author} {\bibfnamefont {A.~T.}\ \bibnamefont
  {Hanbicki}}, \bibinfo {author} {\bibfnamefont {H.-J.}\ \bibnamefont
  {Chuang}}, \bibinfo {author} {\bibfnamefont {M.~R.}\ \bibnamefont
  {Rosenberger}}, \bibinfo {author} {\bibfnamefont {C.~S.}\ \bibnamefont
  {Hellberg}}, \bibinfo {author} {\bibfnamefont {S.~V.}\ \bibnamefont
  {Sivaram}}, \bibinfo {author} {\bibfnamefont {K.~M.}\ \bibnamefont
  {McCreary}}, \bibinfo {author} {\bibfnamefont {I.~I.}\ \bibnamefont
  {Mazin}},\ and\ \bibinfo {author} {\bibfnamefont {B.~T.}\ \bibnamefont
  {Jonker}},\ }\bibfield  {title} {\bibinfo {title} {Double indirect interlayer
  exciton in a {MoSe}$_2$/{WSe}$_2$ van der {W}aals heterostructure},\
  }\href@noop {} {\bibfield  {journal} {\bibinfo  {journal} {ACS Nano}\
  }\textbf {\bibinfo {volume} {12}},\ \bibinfo {pages} {4719} (\bibinfo {year}
  {2018})}\BibitemShut {NoStop}%
\bibitem [{\citenamefont {Joe}\ \emph {et~al.}(2021)\citenamefont {Joe},
  \citenamefont {Jauregui}, \citenamefont {Pistunova}, \citenamefont
  {Valdivia}, \citenamefont {Lu}, \citenamefont {Wild}, \citenamefont {Scuri},
  \citenamefont {De~Greve}, \citenamefont {Gelly}, \citenamefont {Zhou},
  \citenamefont {Sung}, \citenamefont {Sushko}, \citenamefont {Taniguchi},
  \citenamefont {Watanabe}, \citenamefont {Smirnov}, \citenamefont {Lukin},
  \citenamefont {Park},\ and\ \citenamefont {Kim}}]{joe2021electrically}%
  \BibitemOpen
  \bibfield  {author} {\bibinfo {author} {\bibfnamefont {A.~Y.}\ \bibnamefont
  {Joe}}, \bibinfo {author} {\bibfnamefont {L.~A.}\ \bibnamefont {Jauregui}},
  \bibinfo {author} {\bibfnamefont {K.}~\bibnamefont {Pistunova}}, \bibinfo
  {author} {\bibfnamefont {A.~M.~M.}\ \bibnamefont {Valdivia}}, \bibinfo
  {author} {\bibfnamefont {Z.}~\bibnamefont {Lu}}, \bibinfo {author}
  {\bibfnamefont {D.~S.}\ \bibnamefont {Wild}}, \bibinfo {author}
  {\bibfnamefont {G.}~\bibnamefont {Scuri}}, \bibinfo {author} {\bibfnamefont
  {K.}~\bibnamefont {De~Greve}}, \bibinfo {author} {\bibfnamefont {R.~J.}\
  \bibnamefont {Gelly}}, \bibinfo {author} {\bibfnamefont {Y.}~\bibnamefont
  {Zhou}}, \bibinfo {author} {\bibfnamefont {J.}~\bibnamefont {Sung}}, \bibinfo
  {author} {\bibfnamefont {A.}~\bibnamefont {Sushko}}, \bibinfo {author}
  {\bibfnamefont {T.}~\bibnamefont {Taniguchi}}, \bibinfo {author}
  {\bibfnamefont {K.}~\bibnamefont {Watanabe}}, \bibinfo {author}
  {\bibfnamefont {D.}~\bibnamefont {Smirnov}}, \bibinfo {author} {\bibfnamefont
  {M.~D.}\ \bibnamefont {Lukin}}, \bibinfo {author} {\bibfnamefont
  {H.}~\bibnamefont {Park}},\ and\ \bibinfo {author} {\bibfnamefont
  {P.}~\bibnamefont {Kim}},\ }\bibfield  {title} {\bibinfo {title}
  {Electrically controlled emission from singlet and triplet exciton species in
  atomically thin light-emitting diodes},\ }\href@noop {} {\bibfield  {journal}
  {\bibinfo  {journal} {Phys. Rev. B}\ }\textbf {\bibinfo {volume} {103}},\
  \bibinfo {pages} {L161411} (\bibinfo {year} {2021})}\BibitemShut {NoStop}%
\bibitem [{\citenamefont {Dominguez}\ \emph {et~al.}(2014)\citenamefont
  {Dominguez}, \citenamefont {Alharbi}, \citenamefont {Alhusain}, \citenamefont
  {Bernussi},\ and\ \citenamefont {Peralta}}]{dominguez2014fourier}%
  \BibitemOpen
  \bibfield  {author} {\bibinfo {author} {\bibfnamefont {D.}~\bibnamefont
  {Dominguez}}, \bibinfo {author} {\bibfnamefont {N.}~\bibnamefont {Alharbi}},
  \bibinfo {author} {\bibfnamefont {M.}~\bibnamefont {Alhusain}}, \bibinfo
  {author} {\bibfnamefont {A.~A.}\ \bibnamefont {Bernussi}},\ and\ \bibinfo
  {author} {\bibfnamefont {L.~G.~d.}\ \bibnamefont {Peralta}},\ }\bibfield
  {title} {\bibinfo {title} {Fourier plane imaging microscopy},\ }\href@noop {}
  {\bibfield  {journal} {\bibinfo  {journal} {J. Appl. Phys.}\ }\textbf
  {\bibinfo {volume} {116}},\ \bibinfo {pages} {103102} (\bibinfo {year}
  {2014})}\BibitemShut {NoStop}%
\bibitem [{\citenamefont {Li}\ \emph {et~al.}(2019)\citenamefont {Li},
  \citenamefont {Wang}, \citenamefont {Lu}, \citenamefont {Khatoniar},
  \citenamefont {Lian}, \citenamefont {Meng}, \citenamefont {Blei},
  \citenamefont {Taniguchi}, \citenamefont {Watanabe}, \citenamefont {McGill},
  \citenamefont {Tongay}, \citenamefont {Menon}, \citenamefont {Smirnov},\ and\
  \citenamefont {Shi}}]{li2019direct}%
  \BibitemOpen
  \bibfield  {author} {\bibinfo {author} {\bibfnamefont {Z.}~\bibnamefont
  {Li}}, \bibinfo {author} {\bibfnamefont {T.}~\bibnamefont {Wang}}, \bibinfo
  {author} {\bibfnamefont {Z.}~\bibnamefont {Lu}}, \bibinfo {author}
  {\bibfnamefont {M.}~\bibnamefont {Khatoniar}}, \bibinfo {author}
  {\bibfnamefont {Z.}~\bibnamefont {Lian}}, \bibinfo {author} {\bibfnamefont
  {Y.}~\bibnamefont {Meng}}, \bibinfo {author} {\bibfnamefont {M.}~\bibnamefont
  {Blei}}, \bibinfo {author} {\bibfnamefont {T.}~\bibnamefont {Taniguchi}},
  \bibinfo {author} {\bibfnamefont {K.}~\bibnamefont {Watanabe}}, \bibinfo
  {author} {\bibfnamefont {S.~A.}\ \bibnamefont {McGill}}, \bibinfo {author}
  {\bibfnamefont {S.}~\bibnamefont {Tongay}}, \bibinfo {author} {\bibfnamefont
  {V.~M.}\ \bibnamefont {Menon}}, \bibinfo {author} {\bibfnamefont
  {D.}~\bibnamefont {Smirnov}},\ and\ \bibinfo {author} {\bibfnamefont {S.-F.}\
  \bibnamefont {Shi}},\ }\bibfield  {title} {\bibinfo {title} {Direct
  observation of gate-tunable dark trions in monolayer {WSe}$_2$},\ }\href@noop
  {} {\bibfield  {journal} {\bibinfo  {journal} {Nano Lett.}\ }\textbf
  {\bibinfo {volume} {19}},\ \bibinfo {pages} {6886} (\bibinfo {year}
  {2019})}\BibitemShut {NoStop}%
\bibitem [{\citenamefont {Nagler}\ \emph {et~al.}(2017)\citenamefont {Nagler},
  \citenamefont {Ballottin}, \citenamefont {Mitioglu}, \citenamefont
  {Mooshammer}, \citenamefont {Paradiso}, \citenamefont {Strunk}, \citenamefont
  {Huber}, \citenamefont {Chernikov}, \citenamefont {Christianen},
  \citenamefont {Schüller},\ and\ \citenamefont {Korn}}]{Nagler2017}%
  \BibitemOpen
  \bibfield  {author} {\bibinfo {author} {\bibfnamefont {P.}~\bibnamefont
  {Nagler}}, \bibinfo {author} {\bibfnamefont {M.~V.}\ \bibnamefont
  {Ballottin}}, \bibinfo {author} {\bibfnamefont {A.~A.}\ \bibnamefont
  {Mitioglu}}, \bibinfo {author} {\bibfnamefont {F.}~\bibnamefont
  {Mooshammer}}, \bibinfo {author} {\bibfnamefont {N.}~\bibnamefont
  {Paradiso}}, \bibinfo {author} {\bibfnamefont {C.}~\bibnamefont {Strunk}},
  \bibinfo {author} {\bibfnamefont {R.}~\bibnamefont {Huber}}, \bibinfo
  {author} {\bibfnamefont {A.}~\bibnamefont {Chernikov}}, \bibinfo {author}
  {\bibfnamefont {P.~C.~M.}\ \bibnamefont {Christianen}}, \bibinfo {author}
  {\bibfnamefont {C.}~\bibnamefont {Schüller}},\ and\ \bibinfo {author}
  {\bibfnamefont {T.}~\bibnamefont {Korn}},\ }\bibfield  {title} {\bibinfo
  {title} {{Giant magnetic splitting inducing near-unity valley polarization in
  van der Waals heterostructures}},\ }\href@noop {} {\bibfield  {journal}
  {\bibinfo  {journal} {Nat. Commun.}\ }\textbf {\bibinfo {volume} {8}},\
  \bibinfo {pages} {1551} (\bibinfo {year} {2017})}\BibitemShut {NoStop}%
\bibitem [{\citenamefont {Wang}\ \emph {et~al.}(2020)\citenamefont {Wang},
  \citenamefont {Miao}, \citenamefont {Li}, \citenamefont {Meng}, \citenamefont
  {Lu}, \citenamefont {Lian}, \citenamefont {Blei}, \citenamefont {Taniguchi},
  \citenamefont {Watanabe}, \citenamefont {Tongay}, \citenamefont {Smirnov},\
  and\ \citenamefont {Shi}}]{Wang2020}%
  \BibitemOpen
  \bibfield  {author} {\bibinfo {author} {\bibfnamefont {T.}~\bibnamefont
  {Wang}}, \bibinfo {author} {\bibfnamefont {S.}~\bibnamefont {Miao}}, \bibinfo
  {author} {\bibfnamefont {Z.}~\bibnamefont {Li}}, \bibinfo {author}
  {\bibfnamefont {Y.}~\bibnamefont {Meng}}, \bibinfo {author} {\bibfnamefont
  {Z.}~\bibnamefont {Lu}}, \bibinfo {author} {\bibfnamefont {Z.}~\bibnamefont
  {Lian}}, \bibinfo {author} {\bibfnamefont {M.}~\bibnamefont {Blei}}, \bibinfo
  {author} {\bibfnamefont {T.}~\bibnamefont {Taniguchi}}, \bibinfo {author}
  {\bibfnamefont {K.}~\bibnamefont {Watanabe}}, \bibinfo {author}
  {\bibfnamefont {S.}~\bibnamefont {Tongay}}, \bibinfo {author} {\bibfnamefont
  {D.}~\bibnamefont {Smirnov}},\ and\ \bibinfo {author} {\bibfnamefont {S.-F.}\
  \bibnamefont {Shi}},\ }\bibfield  {title} {\bibinfo {title} {Giant
  {{Valley}}-{{Zeeman Splitting}} from {{Spin}}-{{Singlet}} and
  {{Spin}}-{{Triplet Interlayer Excitons}} in {{WSe}}{$_2$}/{{MoSe}}{$_2$}
  {{Heterostructure}}},\ }\href@noop {} {\bibfield  {journal} {\bibinfo
  {journal} {Nano Lett.}\ }\textbf {\bibinfo {volume} {20}},\ \bibinfo {pages}
  {694} (\bibinfo {year} {2020})}\BibitemShut {NoStop}%
\bibitem [{\citenamefont {Delhomme}\ \emph {et~al.}(2020)\citenamefont
  {Delhomme}, \citenamefont {Vaclavkova}, \citenamefont {Slobodeniuk},
  \citenamefont {Orlita}, \citenamefont {Potemski}, \citenamefont {Basko},
  \citenamefont {Watanabe}, \citenamefont {Taniguchi}, \citenamefont {Mauro},
  \citenamefont {Barreteau}, \citenamefont {Giannini}, \citenamefont
  {Morpurgo}, \citenamefont {Ubrig},\ and\ \citenamefont
  {Faugeras}}]{Delhomme2020}%
  \BibitemOpen
  \bibfield  {author} {\bibinfo {author} {\bibfnamefont {A.}~\bibnamefont
  {Delhomme}}, \bibinfo {author} {\bibfnamefont {D.}~\bibnamefont
  {Vaclavkova}}, \bibinfo {author} {\bibfnamefont {A.}~\bibnamefont
  {Slobodeniuk}}, \bibinfo {author} {\bibfnamefont {M.}~\bibnamefont {Orlita}},
  \bibinfo {author} {\bibfnamefont {M.}~\bibnamefont {Potemski}}, \bibinfo
  {author} {\bibfnamefont {D.~M.}\ \bibnamefont {Basko}}, \bibinfo {author}
  {\bibfnamefont {K.}~\bibnamefont {Watanabe}}, \bibinfo {author}
  {\bibfnamefont {T.}~\bibnamefont {Taniguchi}}, \bibinfo {author}
  {\bibfnamefont {D.}~\bibnamefont {Mauro}}, \bibinfo {author} {\bibfnamefont
  {C.}~\bibnamefont {Barreteau}}, \bibinfo {author} {\bibfnamefont
  {E.}~\bibnamefont {Giannini}}, \bibinfo {author} {\bibfnamefont {A.~F.}\
  \bibnamefont {Morpurgo}}, \bibinfo {author} {\bibfnamefont {N.}~\bibnamefont
  {Ubrig}},\ and\ \bibinfo {author} {\bibfnamefont {C.}~\bibnamefont
  {Faugeras}},\ }\bibfield  {title} {\bibinfo {title} {Flipping exciton angular
  momentum with chiral phonons in {MoSe}$_2$/{WSe}$_2$ heterobilayers},\
  }\href@noop {} {\bibfield  {journal} {\bibinfo  {journal} {2D Mater.}\
  }\textbf {\bibinfo {volume} {7}},\ \bibinfo {pages} {041002} (\bibinfo {year}
  {2020})}\BibitemShut {NoStop}%
\bibitem [{\citenamefont {Brotons-Gisbert}\ \emph {et~al.}(2020)\citenamefont
  {Brotons-Gisbert}, \citenamefont {Baek}, \citenamefont {Molina-S{\'a}nchez},
  \citenamefont {Campbell}, \citenamefont {Scerri}, \citenamefont {White},
  \citenamefont {Watanabe}, \citenamefont {Taniguchi}, \citenamefont {Bonato},\
  and\ \citenamefont {Gerardot}}]{brotons2020spin}%
  \BibitemOpen
  \bibfield  {author} {\bibinfo {author} {\bibfnamefont {M.}~\bibnamefont
  {Brotons-Gisbert}}, \bibinfo {author} {\bibfnamefont {H.}~\bibnamefont
  {Baek}}, \bibinfo {author} {\bibfnamefont {A.}~\bibnamefont
  {Molina-S{\'a}nchez}}, \bibinfo {author} {\bibfnamefont {A.}~\bibnamefont
  {Campbell}}, \bibinfo {author} {\bibfnamefont {E.}~\bibnamefont {Scerri}},
  \bibinfo {author} {\bibfnamefont {D.}~\bibnamefont {White}}, \bibinfo
  {author} {\bibfnamefont {K.}~\bibnamefont {Watanabe}}, \bibinfo {author}
  {\bibfnamefont {T.}~\bibnamefont {Taniguchi}}, \bibinfo {author}
  {\bibfnamefont {C.}~\bibnamefont {Bonato}},\ and\ \bibinfo {author}
  {\bibfnamefont {B.~D.}\ \bibnamefont {Gerardot}},\ }\bibfield  {title}
  {\bibinfo {title} {Spin--layer locking of interlayer excitons trapped in
  moir{\'e} potentials},\ }\href@noop {} {\bibfield  {journal} {\bibinfo
  {journal} {Nat. Mater.}\ }\textbf {\bibinfo {volume} {19}},\ \bibinfo {pages}
  {630} (\bibinfo {year} {2020})}\BibitemShut {NoStop}%
\bibitem [{\citenamefont {Robert}\ \emph {et~al.}(2017)\citenamefont {Robert},
  \citenamefont {Amand}, \citenamefont {Cadiz}, \citenamefont {Lagarde},
  \citenamefont {Courtade}, \citenamefont {Manca}, \citenamefont {Taniguchi},
  \citenamefont {Watanabe}, \citenamefont {Urbaszek},\ and\ \citenamefont
  {Marie}}]{robert2017fine}%
  \BibitemOpen
  \bibfield  {author} {\bibinfo {author} {\bibfnamefont {C.}~\bibnamefont
  {Robert}}, \bibinfo {author} {\bibfnamefont {T.}~\bibnamefont {Amand}},
  \bibinfo {author} {\bibfnamefont {F.}~\bibnamefont {Cadiz}}, \bibinfo
  {author} {\bibfnamefont {D.}~\bibnamefont {Lagarde}}, \bibinfo {author}
  {\bibfnamefont {E.}~\bibnamefont {Courtade}}, \bibinfo {author}
  {\bibfnamefont {M.}~\bibnamefont {Manca}}, \bibinfo {author} {\bibfnamefont
  {T.}~\bibnamefont {Taniguchi}}, \bibinfo {author} {\bibfnamefont
  {K.}~\bibnamefont {Watanabe}}, \bibinfo {author} {\bibfnamefont
  {B.}~\bibnamefont {Urbaszek}},\ and\ \bibinfo {author} {\bibfnamefont
  {X.}~\bibnamefont {Marie}},\ }\bibfield  {title} {\bibinfo {title} {Fine
  structure and lifetime of dark excitons in transition metal dichalcogenide
  monolayers},\ }\href@noop {} {\bibfield  {journal} {\bibinfo  {journal}
  {Phys. Rev. B}\ }\textbf {\bibinfo {volume} {96}},\ \bibinfo {pages} {155423}
  (\bibinfo {year} {2017})}\BibitemShut {NoStop}%
\bibitem [{\citenamefont {Chaves}\ \emph {et~al.}(2022)\citenamefont {Chaves},
  \citenamefont {Covaci}, \citenamefont {Peeters},\ and\ \citenamefont
  {Milošević}}]{Chaves2022}%
  \BibitemOpen
  \bibfield  {author} {\bibinfo {author} {\bibfnamefont {A.}~\bibnamefont
  {Chaves}}, \bibinfo {author} {\bibfnamefont {L.}~\bibnamefont {Covaci}},
  \bibinfo {author} {\bibfnamefont {F.~M.}\ \bibnamefont {Peeters}},\ and\
  \bibinfo {author} {\bibfnamefont {M.~V.}\ \bibnamefont {Milošević}},\
  }\bibfield  {title} {\bibinfo {title} {{Topologically protected moir\'e
  exciton at a twist-boundary in a van der Waals heterostructure}},\
  }\href@noop {} {\bibfield  {journal} {\bibinfo  {journal} {2D Mater.}\
  }\textbf {\bibinfo {volume} {9}},\ \bibinfo {pages} {025012} (\bibinfo {year}
  {2022})}\BibitemShut {NoStop}%
\end{thebibliography}

\begin{thebibliography}{11}%
\makeatletter
\providecommand \@ifxundefined [1]{%
 \@ifx{#1\undefined}
}%
\providecommand \@ifnum [1]{%
 \ifnum #1\expandafter \@firstoftwo
 \else \expandafter \@secondoftwo
 \fi
}%
\providecommand \@ifx [1]{%
 \ifx #1\expandafter \@firstoftwo
 \else \expandafter \@secondoftwo
 \fi
}%
\providecommand \natexlab [1]{#1}%
\providecommand \enquote  [1]{``#1''}%
\providecommand \bibnamefont  [1]{#1}%
\providecommand \bibfnamefont [1]{#1}%
\providecommand \citenamefont [1]{#1}%
\providecommand \href@noop [0]{\@secondoftwo}%
\providecommand \href [0]{\begingroup \@sanitize@url \@href}%
\providecommand \@href[1]{\@@startlink{#1}\@@href}%
\providecommand \@@href[1]{\endgroup#1\@@endlink}%
\providecommand \@sanitize@url [0]{\catcode `\\12\catcode `\$12\catcode
  `\&12\catcode `\#12\catcode `\^12\catcode `\_12\catcode `\%12\relax}%
\providecommand \@@startlink[1]{}%
\providecommand \@@endlink[0]{}%
\providecommand \url  [0]{\begingroup\@sanitize@url \@url }%
\providecommand \@url [1]{\endgroup\@href {#1}{\urlprefix }}%
\providecommand \urlprefix  [0]{URL }%
\providecommand \Eprint [0]{\href }%
\providecommand \doibase [0]{https://doi.org/}%
\providecommand \selectlanguage [0]{\@gobble}%
\providecommand \bibinfo  [0]{\@secondoftwo}%
\providecommand \bibfield  [0]{\@secondoftwo}%
\providecommand \translation [1]{[#1]}%
\providecommand \BibitemOpen [0]{}%
\providecommand \bibitemStop [0]{}%
\providecommand \bibitemNoStop [0]{.\EOS\space}%
\providecommand \EOS [0]{\spacefactor3000\relax}%
\providecommand \BibitemShut  [1]{\csname bibitem#1\endcsname}%
\let\auto@bib@innerbib\@empty
\bibitem [{\citenamefont {Bilgin}\ \emph {et~al.}(2015)\citenamefont {Bilgin},
  \citenamefont {Liu}, \citenamefont {Vargas}, \citenamefont {Winchester},
  \citenamefont {Man}, \citenamefont {Upmanyu}, \citenamefont {Dani},
  \citenamefont {Gupta}, \citenamefont {Talapatra}, \citenamefont {Mohite},\
  and\ \citenamefont {Kar}}]{bilgin2015chemical}%
  \BibitemOpen
  \bibfield  {author} {\bibinfo {author} {\bibfnamefont {I.}~\bibnamefont
  {Bilgin}}, \bibinfo {author} {\bibfnamefont {F.}~\bibnamefont {Liu}},
  \bibinfo {author} {\bibfnamefont {A.}~\bibnamefont {Vargas}}, \bibinfo
  {author} {\bibfnamefont {A.}~\bibnamefont {Winchester}}, \bibinfo {author}
  {\bibfnamefont {M.~K.}\ \bibnamefont {Man}}, \bibinfo {author} {\bibfnamefont
  {M.}~\bibnamefont {Upmanyu}}, \bibinfo {author} {\bibfnamefont {K.~M.}\
  \bibnamefont {Dani}}, \bibinfo {author} {\bibfnamefont {G.}~\bibnamefont
  {Gupta}}, \bibinfo {author} {\bibfnamefont {S.}~\bibnamefont {Talapatra}},
  \bibinfo {author} {\bibfnamefont {A.~D.}\ \bibnamefont {Mohite}},\ and\
  \bibinfo {author} {\bibfnamefont {S.}~\bibnamefont {Kar}},\ }\bibfield
  {title} {\bibinfo {title} {Chemical vapor deposition synthesized atomically
  thin molybdenum disulfide with optoelectronic-grade crystalline quality},\
  }\href@noop {} {\bibfield  {journal} {\bibinfo  {journal} {ACS Nano}\
  }\textbf {\bibinfo {volume} {9}},\ \bibinfo {pages} {8822} (\bibinfo {year}
  {2015})}\BibitemShut {NoStop}%
\bibitem [{\citenamefont {Pizzocchero}\ \emph {et~al.}(2016)\citenamefont
  {Pizzocchero}, \citenamefont {Gammelgaard}, \citenamefont {Jessen},
  \citenamefont {Caridad}, \citenamefont {Wang}, \citenamefont {Hone},
  \citenamefont {B{\o}ggild},\ and\ \citenamefont
  {Booth}}]{pizzocchero2016hot}%
  \BibitemOpen
  \bibfield  {author} {\bibinfo {author} {\bibfnamefont {F.}~\bibnamefont
  {Pizzocchero}}, \bibinfo {author} {\bibfnamefont {L.}~\bibnamefont
  {Gammelgaard}}, \bibinfo {author} {\bibfnamefont {B.~S.}\ \bibnamefont
  {Jessen}}, \bibinfo {author} {\bibfnamefont {J.~M.}\ \bibnamefont {Caridad}},
  \bibinfo {author} {\bibfnamefont {L.}~\bibnamefont {Wang}}, \bibinfo {author}
  {\bibfnamefont {J.}~\bibnamefont {Hone}}, \bibinfo {author} {\bibfnamefont
  {P.}~\bibnamefont {B{\o}ggild}},\ and\ \bibinfo {author} {\bibfnamefont
  {T.~J.}\ \bibnamefont {Booth}},\ }\bibfield  {title} {\bibinfo {title} {The
  hot pick-up technique for batch assembly of van der {W}aals
  heterostructures},\ }\href@noop {} {\bibfield  {journal} {\bibinfo  {journal}
  {Nat. Commun.}\ }\textbf {\bibinfo {volume} {7}},\ \bibinfo {pages} {11894}
  (\bibinfo {year} {2016})}\BibitemShut {NoStop}%
\bibitem [{\citenamefont {Purdie}\ \emph {et~al.}(2018)\citenamefont {Purdie},
  \citenamefont {Pugno}, \citenamefont {Taniguchi}, \citenamefont {Watanabe},
  \citenamefont {Ferrari},\ and\ \citenamefont
  {Lombardo}}]{purdie2018cleaning}%
  \BibitemOpen
  \bibfield  {author} {\bibinfo {author} {\bibfnamefont {D.}~\bibnamefont
  {Purdie}}, \bibinfo {author} {\bibfnamefont {N.}~\bibnamefont {Pugno}},
  \bibinfo {author} {\bibfnamefont {T.}~\bibnamefont {Taniguchi}}, \bibinfo
  {author} {\bibfnamefont {K.}~\bibnamefont {Watanabe}}, \bibinfo {author}
  {\bibfnamefont {A.}~\bibnamefont {Ferrari}},\ and\ \bibinfo {author}
  {\bibfnamefont {A.}~\bibnamefont {Lombardo}},\ }\bibfield  {title} {\bibinfo
  {title} {Cleaning interfaces in layered materials heterostructures},\
  }\href@noop {} {\bibfield  {journal} {\bibinfo  {journal} {Nat. Commun.}\
  }\textbf {\bibinfo {volume} {9}},\ \bibinfo {pages} {5387} (\bibinfo {year}
  {2018})}\BibitemShut {NoStop}%
\bibitem [{\citenamefont {Ashida}\ \emph {et~al.}(2015)\citenamefont {Ashida},
  \citenamefont {Kajino}, \citenamefont {Kutsuma}, \citenamefont {Ohtani},\
  and\ \citenamefont {Kaneko}}]{ashida2015crystallographic}%
  \BibitemOpen
  \bibfield  {author} {\bibinfo {author} {\bibfnamefont {K.}~\bibnamefont
  {Ashida}}, \bibinfo {author} {\bibfnamefont {T.}~\bibnamefont {Kajino}},
  \bibinfo {author} {\bibfnamefont {Y.}~\bibnamefont {Kutsuma}}, \bibinfo
  {author} {\bibfnamefont {N.}~\bibnamefont {Ohtani}},\ and\ \bibinfo {author}
  {\bibfnamefont {T.}~\bibnamefont {Kaneko}},\ }\bibfield  {title} {\bibinfo
  {title} {Crystallographic orientation dependence of {SEM} contrast revealed
  by {SiC} polytypes},\ }\href@noop {} {\bibfield  {journal} {\bibinfo
  {journal} {J. Vac. Sci. Technol. B}\ }\textbf {\bibinfo {volume} {33}},\
  \bibinfo {pages} {04E104} (\bibinfo {year} {2015})}\BibitemShut {NoStop}%
\bibitem [{\citenamefont {Andersen}\ \emph {et~al.}(2021)\citenamefont
  {Andersen}, \citenamefont {Scuri}, \citenamefont {Sushko}, \citenamefont
  {De~Greve}, \citenamefont {Sung}, \citenamefont {Zhou}, \citenamefont {Wild},
  \citenamefont {Gelly}, \citenamefont {Heo}, \citenamefont {B{\'e}rub{\'e}},
  \citenamefont {Joe}, \citenamefont {Jauregui}, \citenamefont {Watanabe},
  \citenamefont {Taniguchi}, \citenamefont {Kim}, \citenamefont {Park},\ and\
  \citenamefont {Lukin}}]{andersen2021excitons}%
  \BibitemOpen
  \bibfield  {author} {\bibinfo {author} {\bibfnamefont {T.~I.}\ \bibnamefont
  {Andersen}}, \bibinfo {author} {\bibfnamefont {G.}~\bibnamefont {Scuri}},
  \bibinfo {author} {\bibfnamefont {A.}~\bibnamefont {Sushko}}, \bibinfo
  {author} {\bibfnamefont {K.}~\bibnamefont {De~Greve}}, \bibinfo {author}
  {\bibfnamefont {J.}~\bibnamefont {Sung}}, \bibinfo {author} {\bibfnamefont
  {Y.}~\bibnamefont {Zhou}}, \bibinfo {author} {\bibfnamefont {D.~S.}\
  \bibnamefont {Wild}}, \bibinfo {author} {\bibfnamefont {R.~J.}\ \bibnamefont
  {Gelly}}, \bibinfo {author} {\bibfnamefont {H.}~\bibnamefont {Heo}}, \bibinfo
  {author} {\bibfnamefont {D.}~\bibnamefont {B{\'e}rub{\'e}}}, \bibinfo
  {author} {\bibfnamefont {A.~Y.}\ \bibnamefont {Joe}}, \bibinfo {author}
  {\bibfnamefont {L.~A.}\ \bibnamefont {Jauregui}}, \bibinfo {author}
  {\bibfnamefont {K.}~\bibnamefont {Watanabe}}, \bibinfo {author}
  {\bibfnamefont {T.}~\bibnamefont {Taniguchi}}, \bibinfo {author}
  {\bibfnamefont {P.}~\bibnamefont {Kim}}, \bibinfo {author} {\bibfnamefont
  {H.}~\bibnamefont {Park}},\ and\ \bibinfo {author} {\bibfnamefont {M.~D.}\
  \bibnamefont {Lukin}},\ }\bibfield  {title} {\bibinfo {title} {Excitons in a
  reconstructed moir{\'e} potential in twisted {WSe}$_2$/{WSe}$_2$
  homobilayers},\ }\href@noop {} {\bibfield  {journal} {\bibinfo  {journal}
  {Nat. Mater.}\ }\textbf {\bibinfo {volume} {20}},\ \bibinfo {pages} {480}
  (\bibinfo {year} {2021})}\BibitemShut {NoStop}%
\bibitem [{\citenamefont {Zhao}\ \emph {et~al.}(2022)\citenamefont {Zhao},
  \citenamefont {Li}, \citenamefont {Huang}, \citenamefont {Rupp},
  \citenamefont {G{\"o}ser}, \citenamefont {Vovk}, \citenamefont {Kruchinin},
  \citenamefont {Watanabe}, \citenamefont {Taniguchi}, \citenamefont {Bilgin},
  \citenamefont {Baimuratov},\ and\ \citenamefont
  {Högele}}]{zhao2022excitons}%
  \BibitemOpen
  \bibfield  {author} {\bibinfo {author} {\bibfnamefont {S.}~\bibnamefont
  {Zhao}}, \bibinfo {author} {\bibfnamefont {Z.}~\bibnamefont {Li}}, \bibinfo
  {author} {\bibfnamefont {X.}~\bibnamefont {Huang}}, \bibinfo {author}
  {\bibfnamefont {A.}~\bibnamefont {Rupp}}, \bibinfo {author} {\bibfnamefont
  {J.}~\bibnamefont {G{\"o}ser}}, \bibinfo {author} {\bibfnamefont {I.~A.}\
  \bibnamefont {Vovk}}, \bibinfo {author} {\bibfnamefont {S.~Y.}\ \bibnamefont
  {Kruchinin}}, \bibinfo {author} {\bibfnamefont {K.}~\bibnamefont {Watanabe}},
  \bibinfo {author} {\bibfnamefont {T.}~\bibnamefont {Taniguchi}}, \bibinfo
  {author} {\bibfnamefont {I.}~\bibnamefont {Bilgin}}, \bibinfo {author}
  {\bibfnamefont {A.~S.}\ \bibnamefont {Baimuratov}},\ and\ \bibinfo {author}
  {\bibfnamefont {A.}~\bibnamefont {Högele}},\ }\bibfield  {title} {\bibinfo
  {title} {Excitons in mesoscopically reconstructed moir{\'e}
  heterostructures},\ }\href@noop {} {\bibfield  {journal} {\bibinfo  {journal}
  {2022, arXiv:2202.11139}\ ,\ \bibinfo {pages} {arXiv preprint
  https://arxiv.org/abs/2202.11139}} (\bibinfo {year} {accessed March 3,
  2022})}\BibitemShut {NoStop}%
\bibitem [{\citenamefont {Carr}\ \emph {et~al.}(2018)\citenamefont {Carr},
  \citenamefont {Massatt}, \citenamefont {Torrisi}, \citenamefont {Cazeaux},
  \citenamefont {Luskin},\ and\ \citenamefont {Kaxiras}}]{carr2018relaxation}%
  \BibitemOpen
  \bibfield  {author} {\bibinfo {author} {\bibfnamefont {S.}~\bibnamefont
  {Carr}}, \bibinfo {author} {\bibfnamefont {D.}~\bibnamefont {Massatt}},
  \bibinfo {author} {\bibfnamefont {S.~B.}\ \bibnamefont {Torrisi}}, \bibinfo
  {author} {\bibfnamefont {P.}~\bibnamefont {Cazeaux}}, \bibinfo {author}
  {\bibfnamefont {M.}~\bibnamefont {Luskin}},\ and\ \bibinfo {author}
  {\bibfnamefont {E.}~\bibnamefont {Kaxiras}},\ }\bibfield  {title} {\bibinfo
  {title} {Relaxation and domain formation in incommensurate two-dimensional
  heterostructures},\ }\href@noop {} {\bibfield  {journal} {\bibinfo  {journal}
  {Phys. Rev. B}\ }\textbf {\bibinfo {volume} {98}},\ \bibinfo {pages} {224102}
  (\bibinfo {year} {2018})}\BibitemShut {NoStop}%
\bibitem [{\citenamefont {Kresse}\ and\ \citenamefont
  {Furthmüller}(1996)}]{Kresse1996}%
  \BibitemOpen
  \bibfield  {author} {\bibinfo {author} {\bibfnamefont {G.}~\bibnamefont
  {Kresse}}\ and\ \bibinfo {author} {\bibfnamefont {J.}~\bibnamefont
  {Furthmüller}},\ }\bibfield  {title} {\bibinfo {title} {Efficient iterative
  schemes for ab initio total-energy calculations using a plane-wave basis
  set},\ }\href@noop {} {\bibfield  {journal} {\bibinfo  {journal} {Phys. Rev.
  B}\ }\textbf {\bibinfo {volume} {54}},\ \bibinfo {pages} {11169} (\bibinfo
  {year} {1996})}\BibitemShut {NoStop}%
\bibitem [{\citenamefont {Furness}\ \emph {et~al.}(2020)\citenamefont
  {Furness}, \citenamefont {Kaplan}, \citenamefont {Ning}, \citenamefont
  {Perdew},\ and\ \citenamefont {Sun}}]{Furness2020}%
  \BibitemOpen
  \bibfield  {author} {\bibinfo {author} {\bibfnamefont {J.~W.}\ \bibnamefont
  {Furness}}, \bibinfo {author} {\bibfnamefont {A.~D.}\ \bibnamefont {Kaplan}},
  \bibinfo {author} {\bibfnamefont {J.}~\bibnamefont {Ning}}, \bibinfo {author}
  {\bibfnamefont {J.~P.}\ \bibnamefont {Perdew}},\ and\ \bibinfo {author}
  {\bibfnamefont {J.}~\bibnamefont {Sun}},\ }\bibfield  {title} {\bibinfo
  {title} {{Accurate and Numerically Efficient ${\mathrm{r}}^{2}\mathrm{SCAN}$
  Meta-Generalized Gradient Approximation}},\ }\href@noop {} {\bibfield
  {journal} {\bibinfo  {journal} {J. Phys. Chem. Lett.}\ }\textbf {\bibinfo
  {volume} {11}},\ \bibinfo {pages} {8208} (\bibinfo {year}
  {2020})}\BibitemShut {NoStop}%
\bibitem [{\citenamefont {Ning}\ \emph {et~al.}(2022)\citenamefont {Ning},
  \citenamefont {Kothakonda}, \citenamefont {Furness}, \citenamefont {Kaplan},
  \citenamefont {Ehlert}, \citenamefont {Brandenburg}, \citenamefont {Perdew},\
  and\ \citenamefont {Sun}}]{Ning2022}%
  \BibitemOpen
  \bibfield  {author} {\bibinfo {author} {\bibfnamefont {J.}~\bibnamefont
  {Ning}}, \bibinfo {author} {\bibfnamefont {M.}~\bibnamefont {Kothakonda}},
  \bibinfo {author} {\bibfnamefont {J.~W.}\ \bibnamefont {Furness}}, \bibinfo
  {author} {\bibfnamefont {A.~D.}\ \bibnamefont {Kaplan}}, \bibinfo {author}
  {\bibfnamefont {S.}~\bibnamefont {Ehlert}}, \bibinfo {author} {\bibfnamefont
  {J.~G.}\ \bibnamefont {Brandenburg}}, \bibinfo {author} {\bibfnamefont
  {J.~P.}\ \bibnamefont {Perdew}},\ and\ \bibinfo {author} {\bibfnamefont
  {J.}~\bibnamefont {Sun}},\ }\bibfield  {title} {\bibinfo {title} {Workhorse
  minimally empirical dispersion-corrected density functional with tests for
  weakly bound systems: ${\mathrm{r}}^{2}\mathrm{SCAN}+\mathrm{rVV}10$},\
  }\href@noop {} {\bibfield  {journal} {\bibinfo  {journal} {Phys. Rev. B}\
  }\textbf {\bibinfo {volume} {106}},\ \bibinfo {pages} {075422} (\bibinfo
  {year} {2022})}\BibitemShut {NoStop}%
\bibitem [{\citenamefont {Blöchl}(1994)}]{Bloechl1994}%
  \BibitemOpen
  \bibfield  {author} {\bibinfo {author} {\bibfnamefont {P.~E.}\ \bibnamefont
  {Blöchl}},\ }\bibfield  {title} {\bibinfo {title} {Projector augmented-wave
  method},\ }\href@noop {} {\bibfield  {journal} {\bibinfo  {journal} {Phys.
  Rev. B}\ }\textbf {\bibinfo {volume} {50}},\ \bibinfo {pages} {17953}
  (\bibinfo {year} {1994})}\BibitemShut {NoStop}%
\end{thebibliography}
\end{document}